\documentclass{IEEEtran}
\usepackage[margin=0.75in]{geometry}
\usepackage{makecell}
\usepackage{booktabs}
\usepackage{amsmath}
\usepackage{amsfonts}
\usepackage{times}
\usepackage{soul}
\usepackage{url}
\usepackage[hidelinks]{hyperref}
\usepackage[utf8]{inputenc}
\usepackage[small]{caption}
\usepackage{graphicx}
\usepackage{amsmath}
\usepackage{amsthm}
\usepackage{booktabs}
\usepackage{algorithm}
\usepackage{algorithmic}
\usepackage{subfig}
\usepackage{array}

\newcolumntype{P}[1]{>{\centering\arraybackslash}p{#1}}
\newcolumntype{M}[1]{>{\centering\arraybackslash}m{#1}}

\title{Analyzing Societal Impact of COVID-19: A Study During the Early Days of the Pandemic}

\begin{document}
%\titlenote{Produces the permission block, and
%  copyright information}
%\subtitle{Extended Abstract}
%\subtitlenote{The full version of the author's guide is available as
%  \texttt{acmart.pdf} document}

%\author{Anonymous Authors}

%\author{Anonymous Authors}

\author{Swaroop Gowdra Shanthakumar, Anand Seetharam, Arti Ramesh
\\ \normalfont Computer Science Department, SUNY Binghamton
\\ \normalfont (sgowdra1, aseethar, artir)@binghamton.edu}
 
% \author{Swaroop Gowdra Shanthakumar, Anand Seetharam, Arti Ramesh
% \affiliations Computer Science Department, SUNY Binghamton
% \email (sgowdra1, aseethar, artir)@binghamton.edu
% }


%\begin{abstract}
%This paper provides a sample of a \LaTeX\ document which conforms,
%somewhat loosely, to the formatting guidelines for
%ACM SIG Proceedings.\footnote{This is an abstract footnote}
%\end{abstract}

%   \begin{CCSXML}
% <concept_id>10003120.10003138.10003140</concept_id>
% <concept_desc>Human-centered computing~Ubiquitous and mobile computing systems and tools</concept_desc>
% <concept_id>10010147.10010257.10010293</concept_id>
% <concept_desc>Computing methodologies~Machine learning approaches</concept_desc>

% \ccsdesc[500]{Human-centered computing~Ubiquitous and mobile computing systems and tools}
% \ccsdesc[500]{Computing methodologies~Machine learning approaches}
%
%

%\keywords{COVID-19, Social Media Analysis, Twitter}

\maketitle

\begin{abstract}
%!TEX root = main.tex
In this paper, we collect and study Twitter communications to understand the societal impact of COVID-19 in the United States during the early days of the pandemic.  With infections soaring rapidly, users took to Twitter asking people to self isolate and quarantine themselves. Users also demanded closure of schools, bars, and restaurants as well as lockdown of cities and states. We methodically collect tweets by identifying and tracking trending COVID-related hashtags. We first manually group the hashtags into six main categories, namely, {\it {\it 1)} General COVID, {\it 2)} Quarantine,  {\it 3)} Panic Buying, {\it 4)} School Closures, {\it 5)} Lockdowns, and {\it 6)} Frustration and Hope}, and study the temporal evolution of tweets in these hashtags.  We conduct a linguistic analysis of words common to all hashtag groups and specific to each hashtag group and identify the chief concerns of people as the pandemic gripped the nation ({\it e.g., exploring bidets as an alternative to toilet paper}). We conduct sentiment analysis and our investigation reveals that people reacted positively to school closures and negatively to the lack of availability of essential goods due to panic buying. We adopt a state-of-the-art semantic role labeling approach to  identify the action words ({\it e.g., fear, test}), which capture the actions people are referring to in the tweets. We then leverage a LSTM-based dependency parsing model to analyze the context of the above-mentioned action words (e.g., verb {\it deal} is accompanied by nouns such as {\it anxiety, stress, and crisis}). Finally, we develop a scalable seeded topic modeling approach to automatically categorize and isolate tweets into hashtag groups and experimentally validate that our topic model provides a grouping similar to our manual grouping. Our study presents a systematic way to construct an aggregated picture of peoples' response to the pandemic and lays the groundwork for future fine-grained linguistic and behavioral analysis.

%Our analysis reveals that COVID-19 gripped the nation during this time as is evidenced by the significant number of trending hashtags.
\end{abstract}

%\begin{IEEEkeywords}
%\end{IEEEkeywords}

\section{Introduction}
\label{sec:intro}
%!TEX root = main.tex

COVID-19 (also known as the novel coronavirus) is a truly global pandemic and has affected humans in all countries of the world. While humanity has seen numerous epidemics including a number of deadly ones over the last two decades (e.g., SARS, MERS, Ebola), the grief and disruption that COVID-19 has already inflicted is incomparable. At the time of writing this paper, COVID-19 is still rapidly spreading around the world and projections for the next few months are grim and extremely disconcerting. The learnings from COVID-19 will also enable humankind to prevent such epidemics from transforming into global pandemics and minimize the socio-economic disruption.

%With no cure in sight and with the chances of COVID-19 reemerging for a second (or multiple) time(s) even after the world manages to contain this first outbreak, it is critical that we understand and analyze the socio-economic disruptions of the first outbreak, so that we are better prepared to handle it in the future. Additionally, with ever-increasing  mobility of humans and goods, it is only prudent to assume that such epidemics are likely to occur in the future. 

In this work, our goal is to analyze the societal impact of COVID-19 in the United States of America during its early days, understand the chain of events that occurred during the spread of the infection, and draw meaningful conclusions so that similar mistakes can be avoided in the future.  Though Twitter data has previously been shown to be biased \cite{morstatter:2017discovering}, Twitter has emerged as the primary media for people to express their opinion especially during this time and our study offers a perspective into the impact as self-disclosed by people in a form that is easily understandable and can be acted upon. We summarize our main contributions below.

We collect 530,206 tweets from Twitter between March $14^{th}$ to March $24^{th}$, a time period when the virus made its first significant inroads into the US and quantitatively demonstrate the disruption and distress experienced by the people.  We group the hashtags into six main categories, namely {\it 1)} General COVID, {\it 2)} Quarantine, {\it 3)} School Closures, {\it 4)}  Panic Buying, {\it 5)} Lockdowns,  and {\it 6)} Frustration and Hope,  to quantitatively and qualitatively understand the chain of events. We observe that general COVID  and quarantine-related messages remain trending throughout the duration of our study. In comparison, we observe calls for closing schools and universities peaking in the middle of March and then reducing when the closures go into effect (e.g., \#closenycschools). We also observe a similar trend with panic buying with essential items particularly toilet paper becoming unavailable in stores (e.g., \#panicbuying, \#toiletpapercrisis). 

%Lockdowns also have a significant number of tweets with calls initially being focused on closure of bars, followed by cities and then states (e.g., \#barsshut, \#seattleshutdown, \#shutdownnyc,  \#shutdownflorida). Tweets in the frustration and hope hashtag group have an overall increasing trend as the struggle with the virus mounts.

%Calls for closures started off with schools, bars, restaurants, and then included entire cities and states  Alongside, panic buying and  hoarding escalated with  We  observe increased calls for social distancing, quarantining, and working from home to limit the spread of the disease (e.g., \#socialdistancingnow, \#workfromhome).  

%With the passage of time, we see an increased fluctuation in emotions with some people expressing their anger at individuals flouting social distancing calls (e.g., \#covidiots), while others rallying people to fight the disease (e.g., \#fightback) and to save workers (e.g., \#saveworkers).

We conduct a linguistic analysis of the tweets in the different hashtag groups and present the words that are representative of each group.  We observe that words such as \textit{family}, \textit{life}, \textit{health}, and \textit{death} are common across hashtag groups. Additionally, for example, if we consider the School Closures category, we observe that unigrams (e.g., \textit{teacher}, \textit{learn}) and bigrams (e.g., \textit{home school}, \textit{kid home}) reflect the most discussed issues.  We also conduct \textit{sentiment analysis} to unearth the overall sentiment of the people. Our investigation reveals that people reacted positively to school closures and negatively to the lack of availability of essential goods due to panic buying. We next adopt a state-of-the-art semantic role labeling approach to identify the action words ({\it e.g., fear, test}) that are uniquely representative in each hashtag group. These action words help understand  peoples' actions in each group. We  leverage a LSTM dependency parsing model to analyze the context of the above-mentioned action words (e.g., verb {\it deal} is accompanied by nouns such as {\it anxiety and stress}). 

Finally, we develop a scalable \textit{seeded topic modeling (seeded LDA)} approach to automatically categorize tweets into specific topics of interest, especially when the topics are rarer in the dataset. We experimentally validate our seeded LDA model and observe that it provides a grouping similar to our manual grouping. Our  study  summarizes the critical public responses surrounding COVID-19, paving the way for future fine-grained linguistic and graph analysis.

%We  observe mentions to \textit{mental health}, a possible consequence of social isolation. We also observe solidarity for essential workers and gratitude towards them (\#saveworkers).

\section{Data and Methods}
\label{sec:data}
%!TEX root = main.tex

In this section, we discuss our methodology for data collection from Twitter to  investigate the societal impact of COVID-19 in the United States   during its early days.  We collect data using the Twitter search API. The results presented in this paper are based on the data collected from March 14 to March 24, 2020. We track the trending COVID related hashtags every day and collect the tweets in those specific hashtags. We repeat this process to collect a total of 530,206 tweets during this time period.

\subsection{Hashtag Categories}
We group the hashtags into six main categories, namely {\it 1)} General COVID, {\it 2)} Quarantine,  {\it 3)} School Closures, {\it 4)}  Panic Buying, {\it 5)} Lockdowns, and  {\it 6)} Frustration and Hope  to quantitatively and qualitatively understand the chain of events. We collect data on per day basis for the different hashtags as and when they become trending. Table \ref{tab_total_tweets} shows the number of tweets in each category, while Table    \ref{tab_hashtags} shows the grouping of some of the representative hashtags by category. We observe that the total number of tweets as grouped by hashtags is 664,476, which is higher than the total number of tweets. This is because tweets can contain multiple hashtags and thus the same tweet can be  grouped into multiple categories. We present some example tweets in Table \ref{tab_tweets} to illustrate the types of communications occurring on Twitter during this period.

\begin{table}[ht]
\caption{Number of Tweets by Category}
\vspace{-2 mm}
\begin{center}
\begin{tabular}{|c|c|}%|M{1.8cm}|M{1.8cm}|M{1.8cm}|M{1.8cm}|M{1.8cm}|M{1.8cm}|M{1.8cm}|}
\hline
\textbf{Category} & \textbf{Number of Tweets} \\
\hline
General Covid &  4,81,398 \\
 \hline
Quarantine &  142,297 \\
\hline
 Lockdowns & 14,709 \\

 \hline
 Frustration and Hope & 13,084 \\
 \hline
Panic Buying & 10,855 \\
 
 \hline
School Closures & 2,133 \\
  \hline
%\multicolumn{5}{l}{$^{\mathrm{a}}$Sample of a Table footnote.}
\end{tabular}
\label{tab_total_tweets}
\end{center}
\vspace{-4mm}
\end{table}

\noindent {\bf 1. General COVID:}  In this category, we group hashtags related to COVID related messages as it is the most discussed topic in conversations. This grouping is done by accumulating hashtags related to COVID-19. 

%such as \#covid19, \#covid, \#corona, \#coronavirus,  \#coronavirus2020, \#coronavirususa, \#outbreak,  \#coronavirusoutbreak, \#covid19outbreak, \#coronaapocalypse, \#cononavirusPandemic, \#californiacoronavirus, \#nyccoronavirus, \#bayareacoronavirus, \#seattlecovid19, \#Floridacoronavirus, \#utahcovid19, \#ohiocoronavirus, \#pandemic  \#coronavirusupdate and \#highriskcovid19  in this category. 

\noindent {\bf 2. Quarantine:} Calls for social distancing and quarantines  flooded Twitter during this outbreak. Communications  centered around quarantines, working from home and flattening the curve to slow the spread of the virus. 

\noindent {\bf 3. School Closures:} In this category, we  collect data related to school closures.   Before states decided to close schools, users on Twitter  demanded the government to shut down public schools and universities. We collect data from a number of  hashtags centered around this call for action.

%such as \#schoolclosures, \#closenypublicschools, \#closenycschools, \#closenyschools, \#suny, \#cuny, \#homeschooling, \#homeschool, \#homeschool2020, \#schoolsclosed, \#noschool, \#closetheschools, \#shutdownschools, \#closethepublicschools and \#schoolsclosing  in this category.

\noindent {\bf 4. Panic Buying:} The spread of the virus also resulted in panic buying  and hoarding. People  rushed to shopping marts and there was a huge panic buying of sanitizers and toilet paper. This panic buying resulted in severe shortage of toilet papers around the middle of March, an issue that remained unresolved till the end of April.

%We grouped hashtags such as \#panicbuying, \#panicshopping, \#panicbuyers, \#toiletpaper, \#notoiletpaper, \#sanitizers, \#handsanitizer, \#toiletpaperpanic, \#toiletpapercrisis, \#toiletpapershortage, \#howmuchtoiletpaper, \#toiletpaperapocolypse, \#washyourhands and \#coronashopping  in this category.

\noindent {\bf 5. Lockdowns:}  With COVID-19 spreading unabated, lock downs of  stores, bars, restaurants, and cities began in many states, resulting in a surge in tweets related to lock downs.

%We grouped hashtags such as \#lockdown, \#shutdown, \#seattleshutdown, \#shutitdown, \#shutdownnyc, \#lalockdown, \#sflockdown, \#bayarealockdown, \#lockdownusa, \#californialockdown, \#vegasshutdown, \#newjerseylockdown, \#barsclosed, \#barsshut, \#californiashutdown, \#newyorklockdown, \#illinoislockdown, \#shutdownflorida, \#floridalockdown, \#ohiolockdown,  \#nyclockdown and \#nycshutdown in this category. The lockdowns took place in cities in different times based on the outbreak in those areas.

%We grouped hashtags such as \#QuarantineLife, \#quarantined, \#staysafestayhome, \#staytheFhome, \#staythefuckhome, \#Quarantinelife, \#socialdistancing, \#socialdistancingnow, \#Quaratineandchill, \#workfromhome, \#homeoffice, \#selfquarantine, \#stayhome, \#stayhomechallenge, \#stayathome, \#stayathomechallenge, \#clubquarantine, \#quarantinecats, \#workingfromhome, \#flattenthecurve and \#flatteningthecurve  in this category.

\noindent {\bf 6. Frustration and Hope:}  Emotions  ran high during these times with people expressing anger and resentment towards those not abiding by social distancing and quarantine rules. Alongside, people also rallied to support workers working hard to keep essential services running. With the beginning of April approaching, many people started to  worry about their next month's rent. 

%We group hashtags such as \#askthemayor, \#coronavirusremedy, \#fuckcovid19, \#canceltherent, \#fightcorona, \#fightback, \#saveyourlife, \#covidiots, \#DrFauci, \#coronafighters, \#letsfightcorona, \#saveworkers, \#savetheday, \#stopthespread, \#whenthisisallover, \#staysafe and \#savelives in this category.

\begin{table*}[ht]
\caption{Hashtags by Category}
\vspace{-2 mm}
\begin{center}
\begin{tabular}{|c|c|}%|M{1.8cm}|M{1.8cm}|M{1.8cm}|M{1.8cm}|M{1.8cm}|M{1.8cm}|M{1.8cm}|}
\hline
\textbf{Category} & \textbf{Hashtags} \\
\hline
General COVID  &  \#covid19, \#COVID19, \#Covid19, \#covid\_19,  \#covid, \#corona, \#coronavirus, \#Coronavirus, \\ 
\hline

Quarantine & \#QuarantineLife, \#quarantined, \#staysafestayhome, \#staytheFhome, \#staythefuckhome, \\
 \hline
 
 School Closures &\#schoolclosures, \#closenypublicschools, \#closenycschools, \#closenyschools, \#suny, \#cuny, \\
\hline
Panic Buying & \#panicbuying, \#PanicBuying, \#panicshopping, \#PanicShopping, \#panicbuyers, \#toiletpaper, \\

\hline
 Lockdowns & \#lockdown, \#Shutdown, \#seattleshutdown, \#ShutItDown, \#shutdownnyc, \#lalockdown, \#sflockdown, \\
 
\hline
 Frustration and Hope &  \#fightback,\#saveyourlife, \#COVIDIOTS, \#COVIDIDIOTS, \#WhenThisIsAllOver, \#coronafighters, \\
\hline
%\multicolumn{5}{l}{$^{\mathrm{a}}$Sample of a Table footnote.}
\end{tabular}
\label{tab_hashtags}
\end{center}
\vspace{-3mm}
\end{table*}

\begin{table*}[ht]
\caption{Example Tweets by Category}
\vspace{-2 mm}
\begin{center}
\begin{tabular}{|c|c|}%|M{1.8cm}|M{1.8cm}|M{1.8cm}|M{1.8cm}|M{1.8cm}|M{1.8cm}|M{1.8cm}|}
\hline
\textbf{Category} & \textbf{Example Tweets} \\
\hline
General COVID &  If this wasn't so f***ing deadly serious I'd be laughing... \#COVID19\\
 \hline
Quarantine &  Extroverts. I get it. You need human interaction to fuel your well being, \\
& but \#StayTheFHome and interact on social media  \\

 \hline
School Closures & If \@ NYCMayorsOffice \@ NYCMayor won't \#closenycpublicschools to protect students and their families, we  \\
& will \#sickout  \#CLOSENYCPUBLICSCHOOLS. Teachers are parents too. We all have family. Keep us all safe. \\

 \hline
Panic Buying & Stop hoarding toilet paper, you morons!   \#ToiletPaperPanic \\

\hline
 Lockdowns & NYC is going to get destroyed. I'm so depressed.\#NYCLockdown \\

 \hline
 Frustration and Hope & When are we going to \#CancelRent in this state? Hundreds of thousands are filing for unemployment and \\
 & can't pay rent. Sure, we can't be evicted, but what's preventing companies from coming after us after this is over? \\

  \hline
%\multicolumn{5}{l}{$^{\mathrm{a}}$Sample of a Table footnote.}
\end{tabular}
\label{tab_tweets}
\end{center}
\vspace{-3mm}
\end{table*}

\subsection {Gaps in Data Collection} Due to the data collection limits imposed by Twitter, we are able to only collect and analyze a portion of the tweets. Though we started collecting data as quickly as we conceived of this project, we were unable to collect data during the first week of March. Though we ran our script to collect data as far back as March 8, because of the way Twitter provides data, we  obtained a limited number of tweets from March 8 to March 13. Additionally, due to the rapidly evolving situation, it is likely that we have inadvertently missed some important hashtags, despite our best efforts. As is the case with most studies based on Twitter data, we also acknowledge the presence of bias in data collection \cite{morstatter:2017discovering}. Having said that, the goal of this study is to provide a panoramic summarized view of the impact of the pandemic on people's lives and aggregate public opinion as expressed by them. Due to the nature of this study, we  are confident that the results presented here help in appreciating the sequence of events that transpired and better prepare ourselves from possible future waves of COVID-19 or another pandemic.

\section{Linguistic Analysis}
\label{sec:experiment}
%!TEX root = main.tex

In this section, we present  observations and results based on our linguistic analysis of the tweets. We study the popularity and temporal evolution of individual hashtags and hashtag groups. We explore the word-usage (i.e, unigram and bigram) frequencies for each hashtag group to understand the main points of discussion.  We then conduct a sentiment analysis to understand the prevailing sentiments in the tweets. We adopt a semantic role labeling approach to identify the action words (i.e., verbs) as well as the corresponding contextual analysis of these action words.  Finally, we develop a scalable seeded LDA based topic model to automatically group tweets and validate its effectiveness with our manual grouping.

\begin{figure}
    \centering
  \subfloat[Number of tweets for the different hashtags]{%
       \includegraphics[scale=0.16]{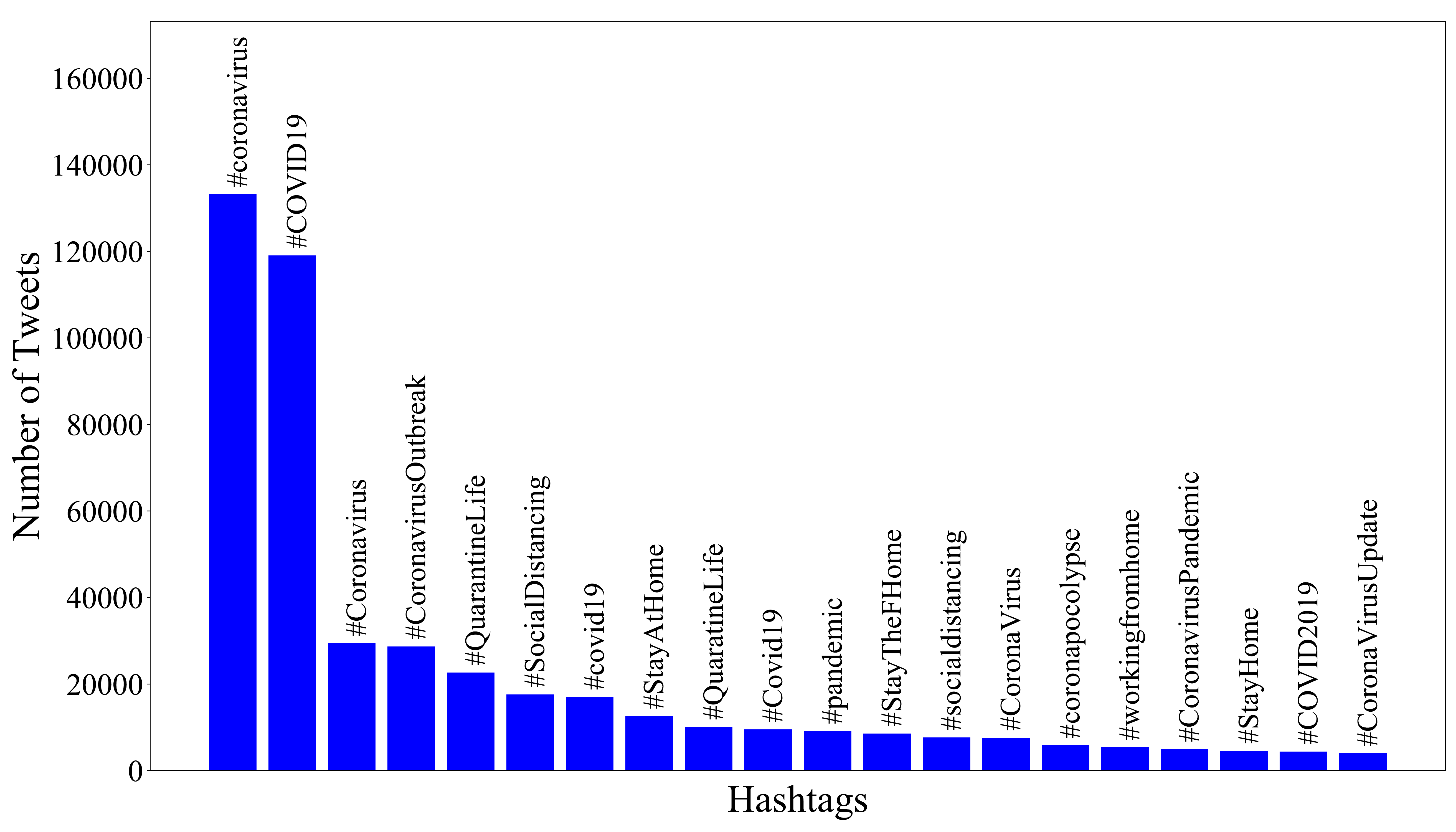}
       \label{fig:hastags_top}}
       \vspace{1mm}
  \subfloat[Most trending hashtags by day]{%
       \includegraphics[scale=0.16]{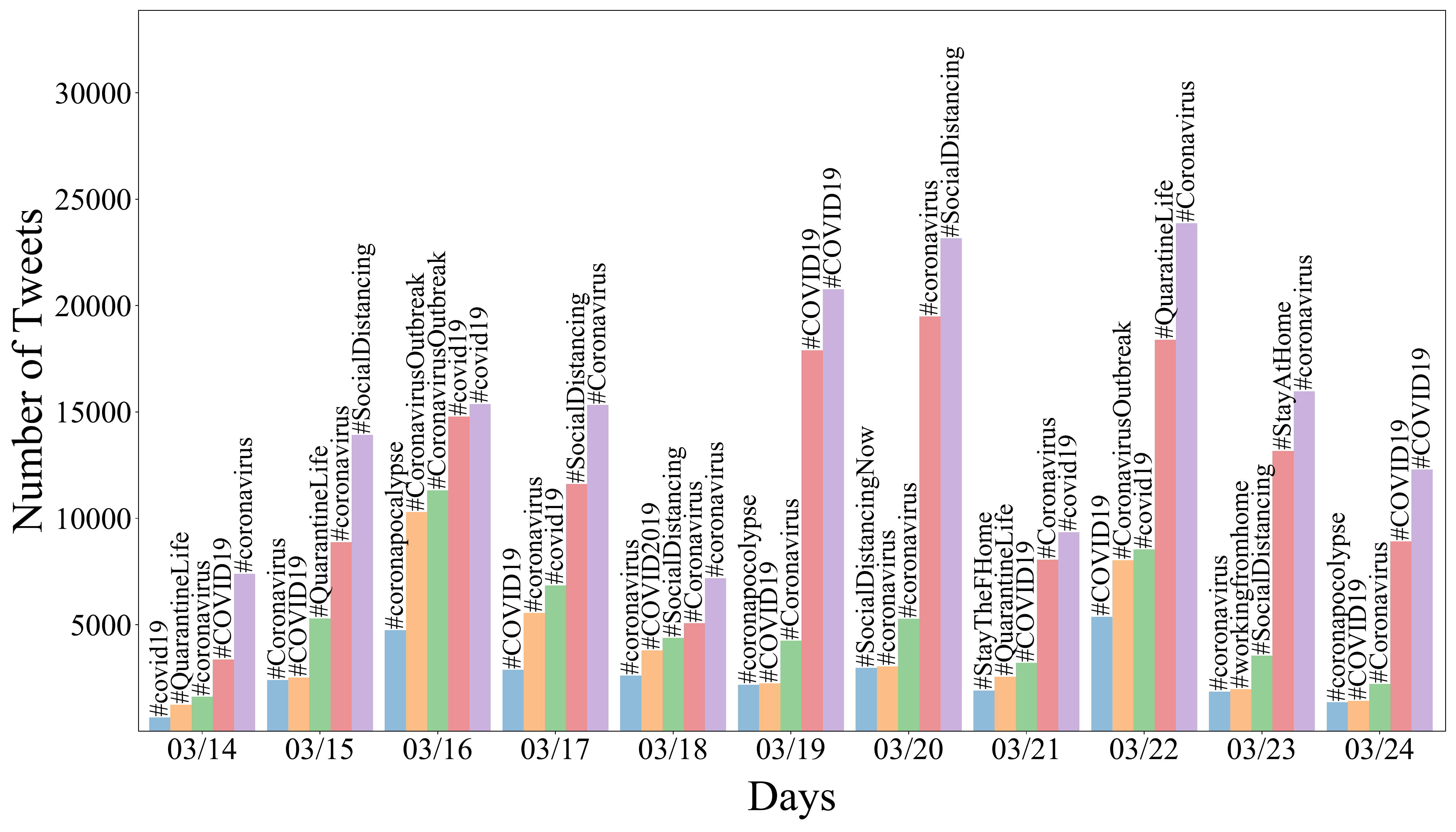}
       \label{fig:hastags_trending}}
       \vspace{1mm}
	\caption{Popularity of different hashtags} 
  \label{fig:hastags} 
    \vspace{-4mm}
\end{figure}

\begin{figure*}[!ht]
    \centering
  \subfloat[General COVID and Quarantine]{%
       \includegraphics[scale=0.15]{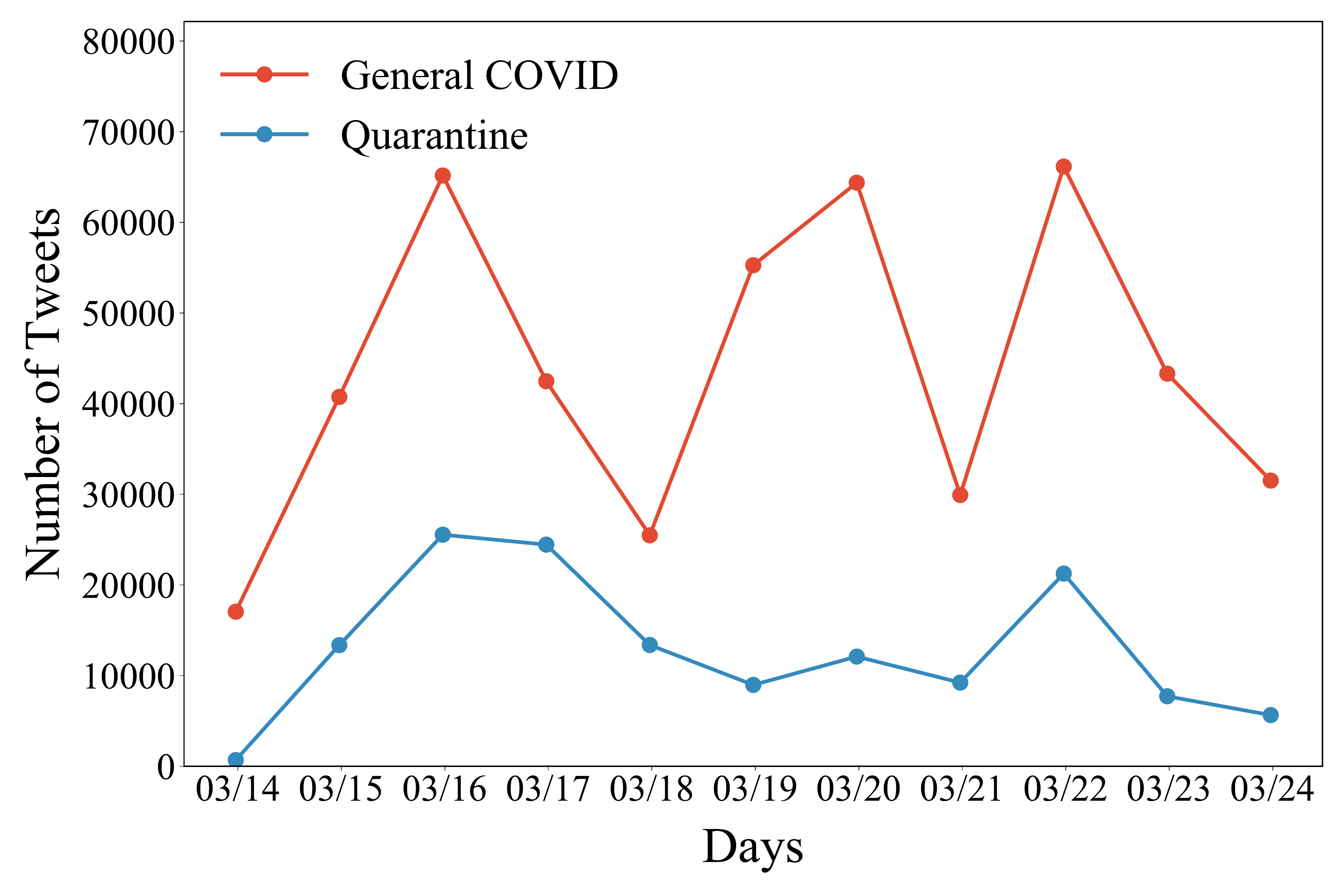}
       \label{fig:covid_quarantine}}
       \vspace{1mm}
  \subfloat[School Closures and Panic Buying]{%
       \includegraphics[scale=0.15]{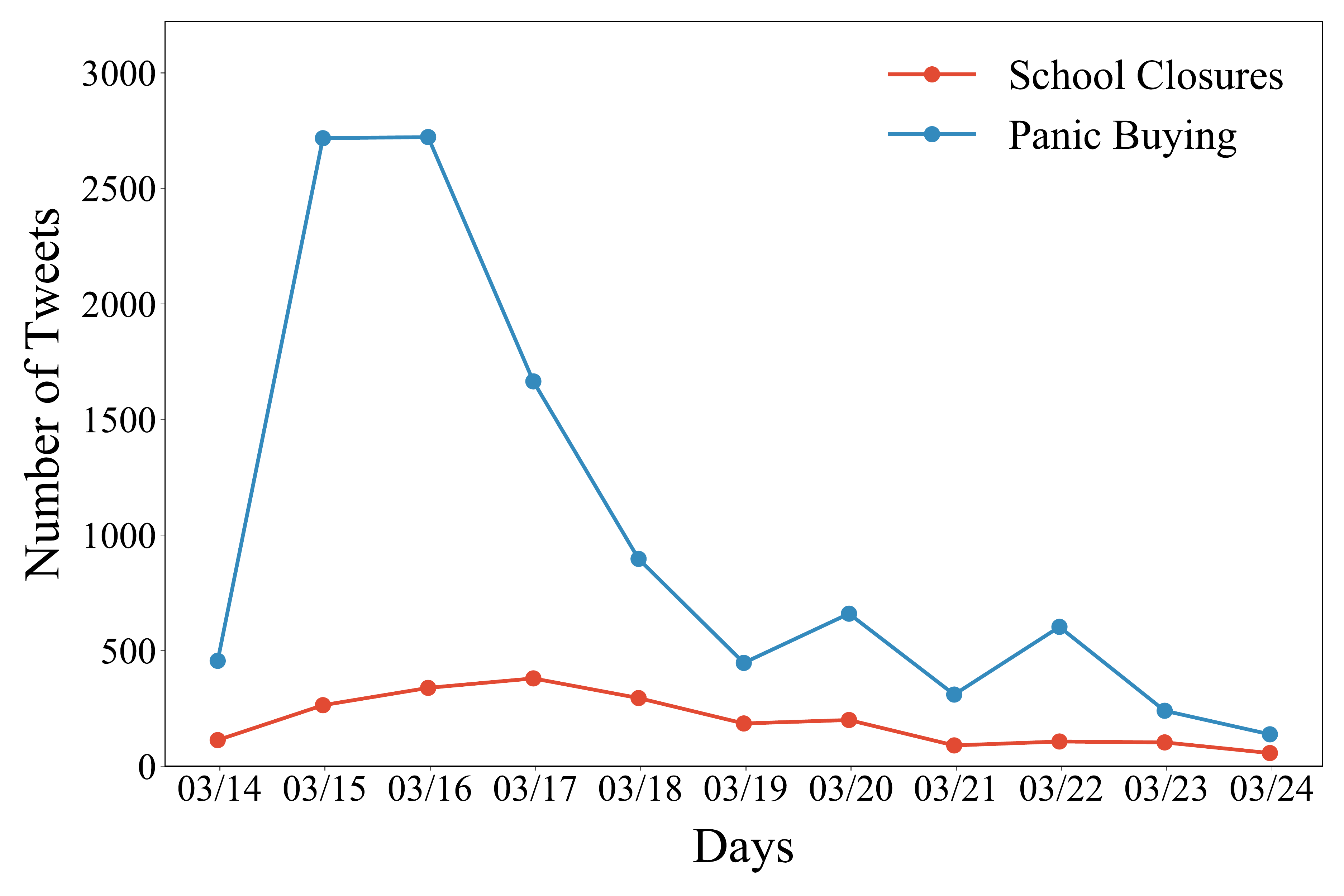}
       \label{fig:panic_school}}
       \vspace{1mm}
  \subfloat[Lockdowns and Frustration and Hope]{%
       \includegraphics[scale=0.15]{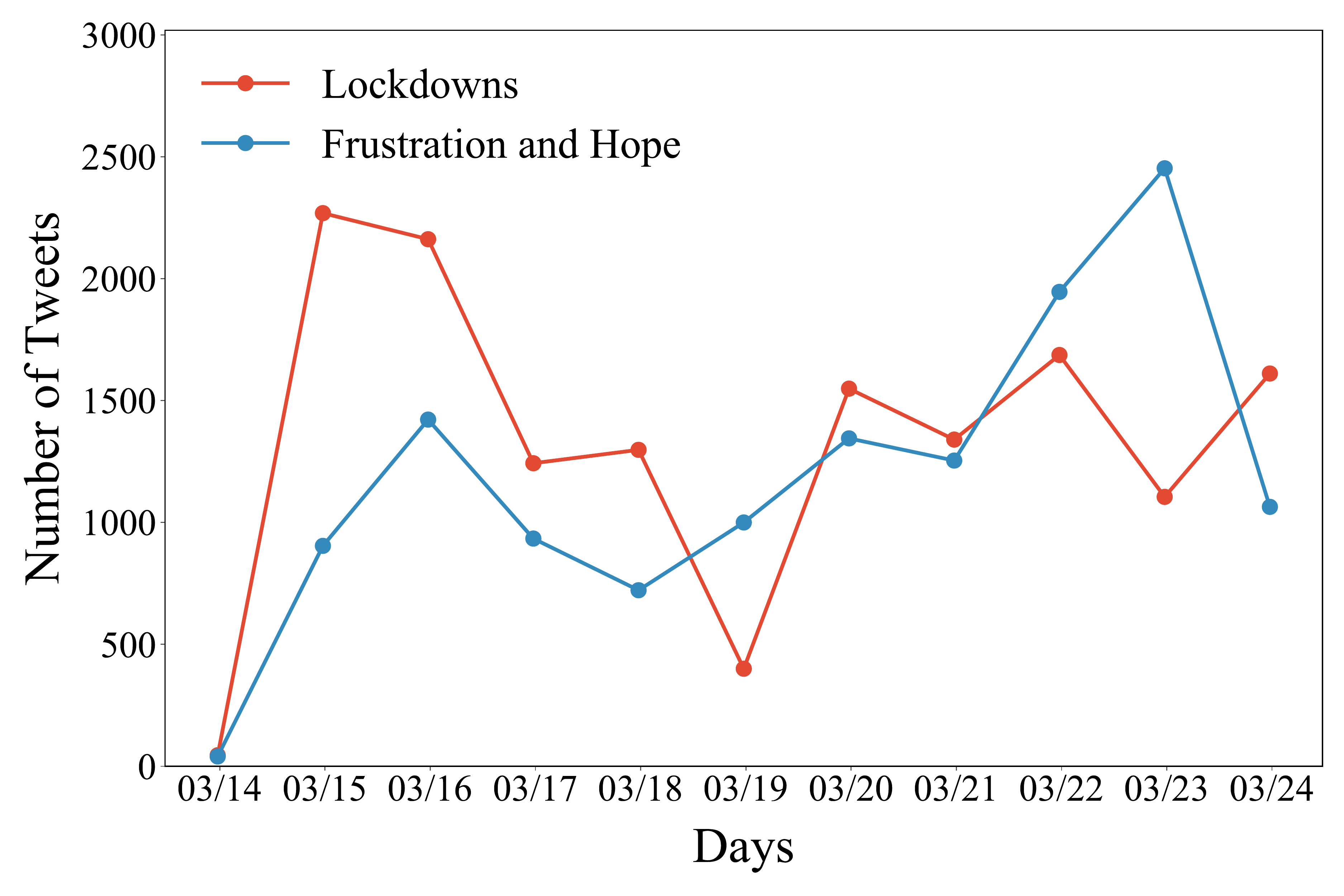}
       \label{fig:lockdown_frustation}}
	\caption{Temporal Evolution of Tweets in Hashtag Groups} 
  \label{fig:mae} 
    \vspace{-4mm}
\end{figure*}

%Our analysis summarizes the critical public responses surrounding COVID-19 and paves the way for more insightful and fine-grained linguistic analysis in the future.

\subsection{Temporal Evolution of Hashtag Groups}
Figure \ref{fig:hastags_top} shows the top 20 hashtags observed in our data. As expected, we see that hashtags corresponding directly to COVID or coronavirus are the most popular hashtags as most communications are centered around them. We observe that hashtags around social isolation, staying at home, and quarantining are also popular. Figure \ref{fig:hastags_trending} shows the most popular hashtags by date. Similar to Figure \ref{fig:hastags_top}, we observe that hashtags related directly to COVID and social distancing trend most on Twitter. The figures and the number of tweets  highlight how the pandemic gripped the United States with its rate of spread.

We investigate the evolution of the number of tweets in various hashtag groups over time. To calculate the number of tweets in each hashtag group, we count the number of mentions of hashtags in that group across all the tweets. If the tweet contains more than one hashtag, it is counted as part of all the hashtags mentioned in it. As the number of tweets for hashtag groups vary significantly, we plot the groups that have similar number of tweets together. Similar to Figure \ref{fig:hastags}, we observe from Figure \ref{fig:covid_quarantine} that the total number of tweets in the General COVID and Quarantine categories are relatively high throughout the time period of the study.

Interestingly, from Figure \ref{fig:panic_school}, we observe that panic buying and calls for school closures peak around the middle of March and then decrease as school closures  and rationing of many essential goods such as toilet paper, cereal, and milk take effect. From Figure \ref{fig:lockdown_frustation}, we see that calls for lock downs related to schools, bars, and cities  peak in the middle of March. With the virus spreading unabated,  we observe intense calls for lock downs of cities and entire states around the beginning of the fourth week of March, resulting in an increased number of tweets in this category. With passage of time, we observe people increasingly expressing their frustration and distress in communications, while some hashtags attempt to inject a more positive outlook.

\begin{figure*}[!ht]
    \centering

  \subfloat[General COVID word frequency]{%
       \includegraphics[scale=0.27]{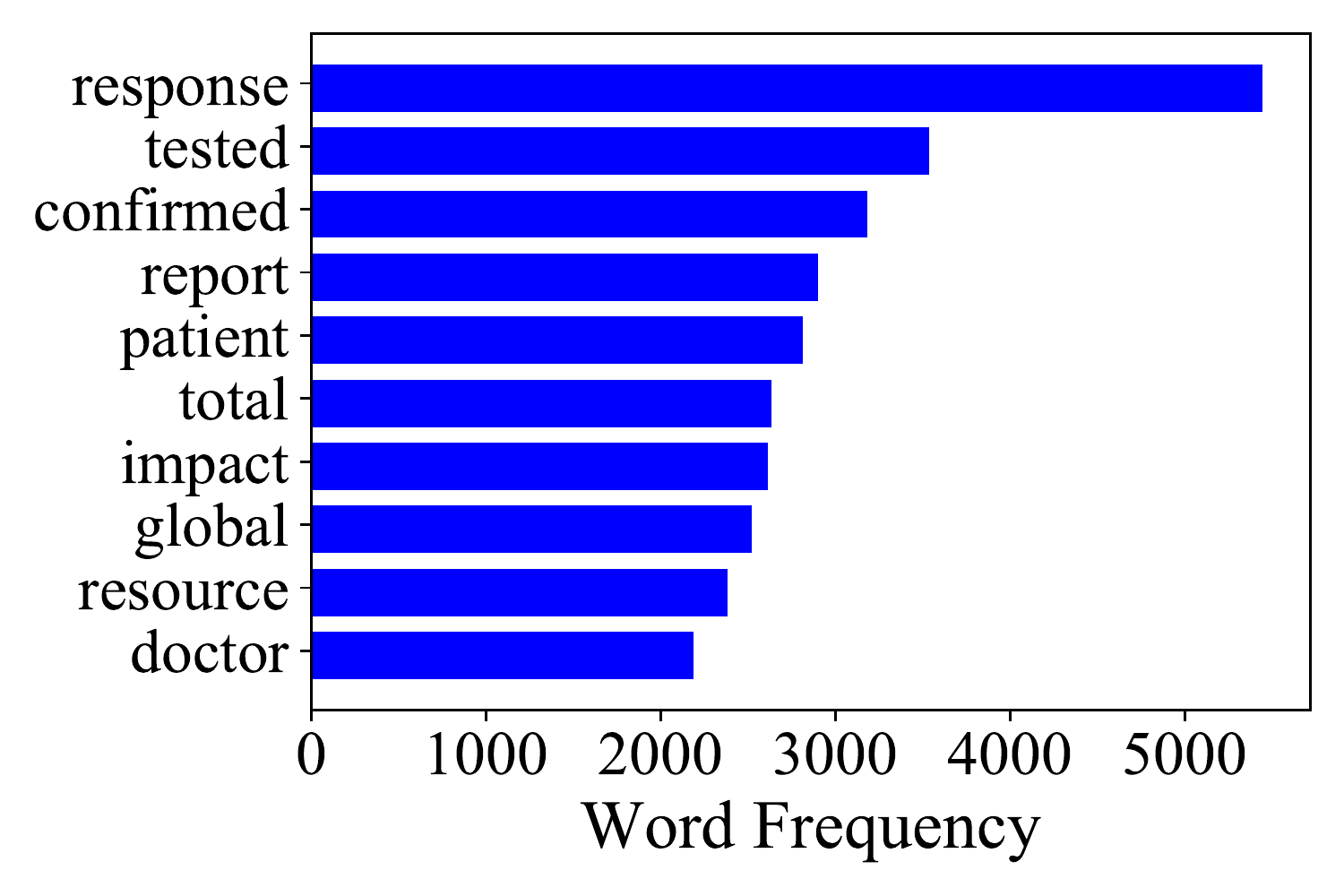}
       \label{fig:covid_word}}
       \vspace{1mm}
  \subfloat[School Closures word frequency]{%
       \includegraphics[scale=0.27]{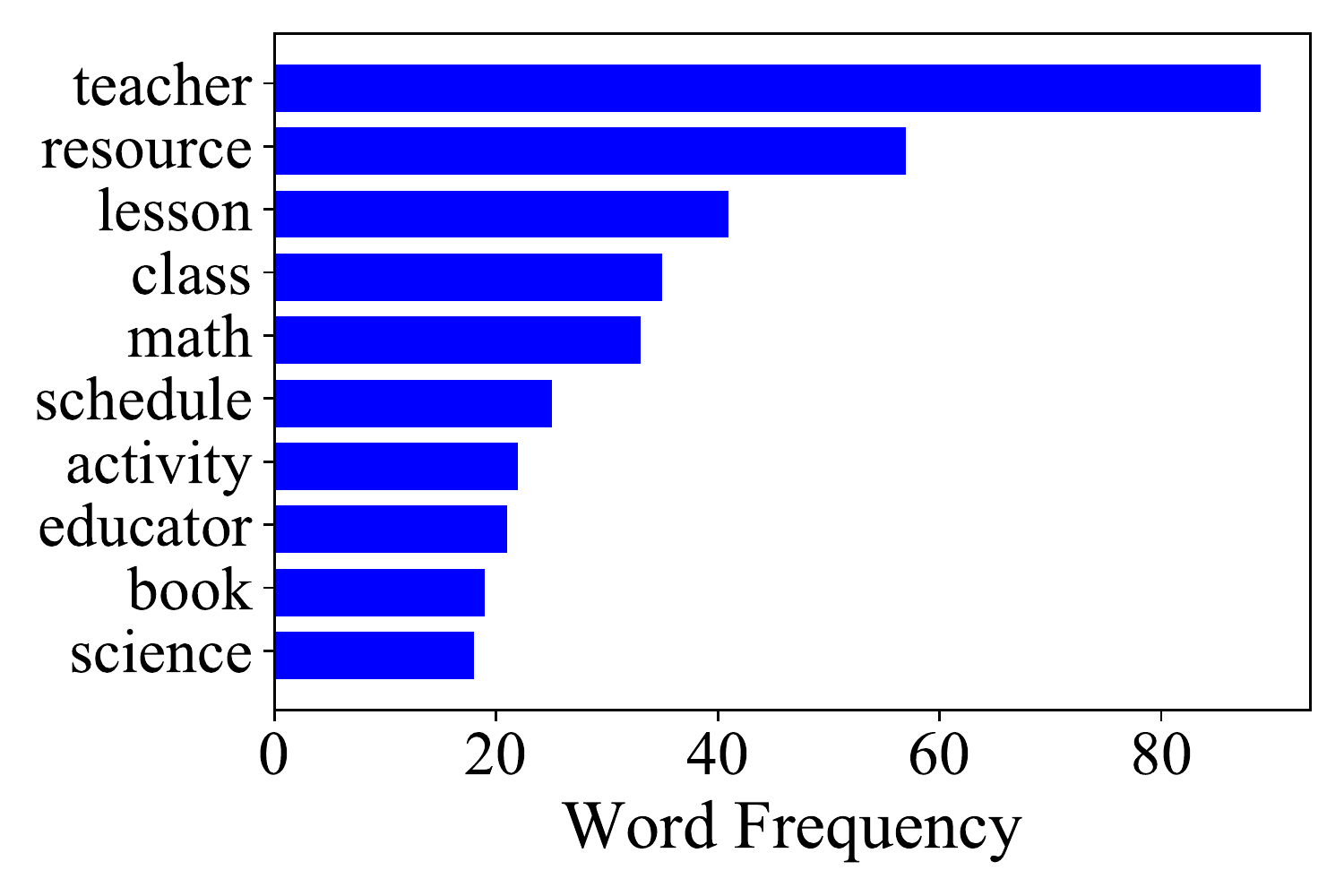}
       \label{fig:school_word}}
       \vspace{1mm}
  \subfloat[Panic Buying word frequency]{%
       \includegraphics[scale=0.27]{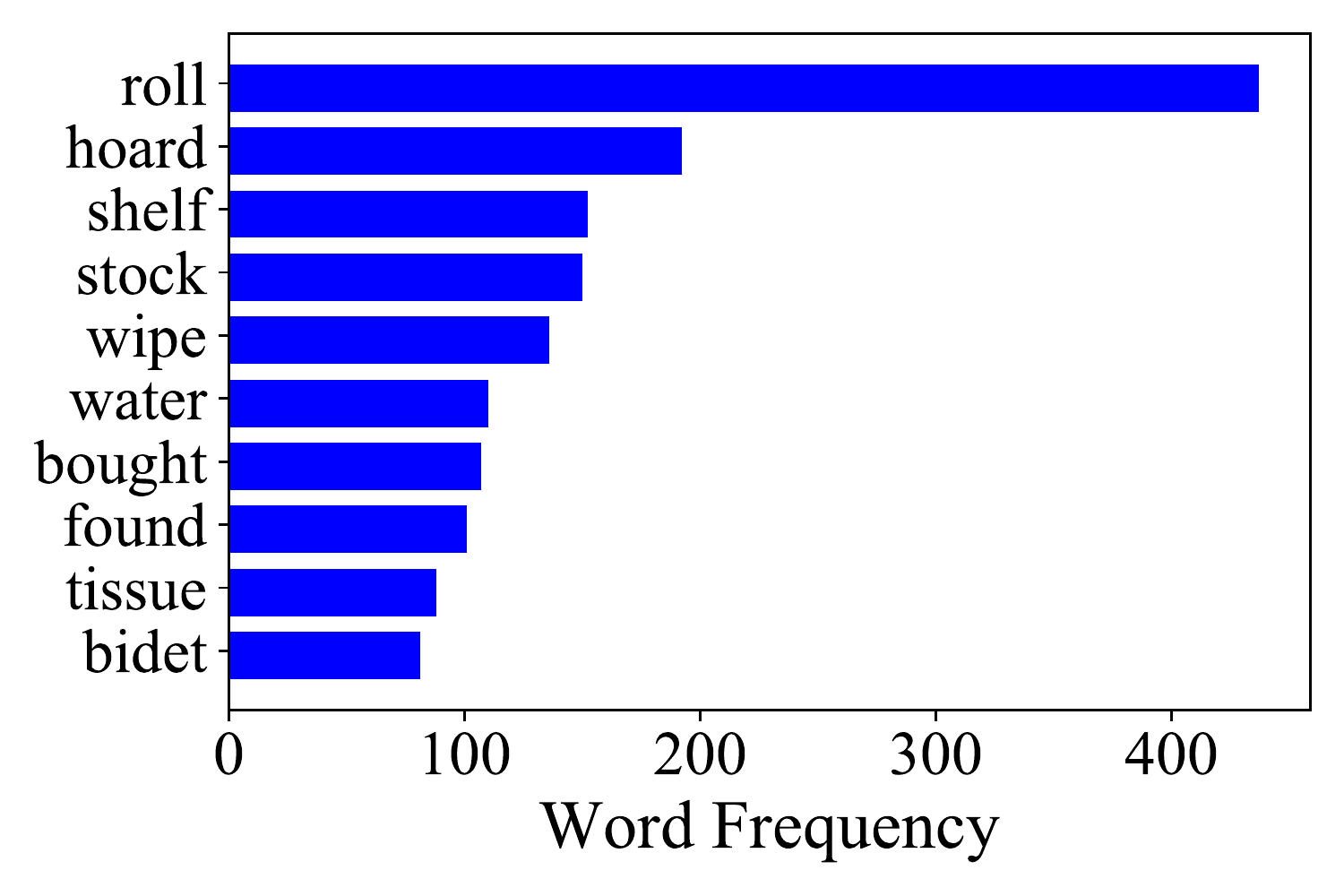}
       \label{fig:panic_word}}
       \vspace{1mm}
  \subfloat[Lockdowns word frequency]{%
       \includegraphics[scale=0.27]{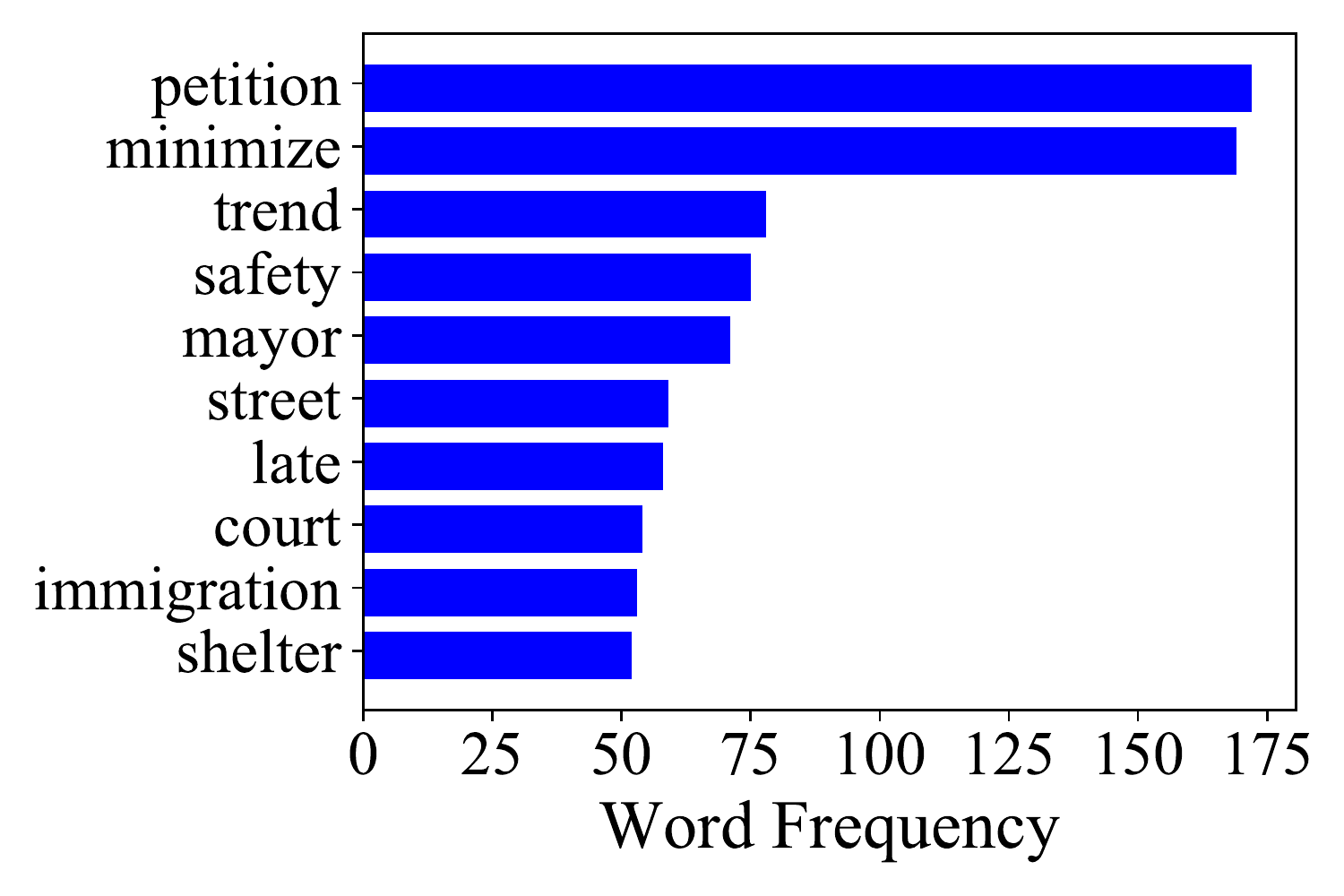}
       \label{fig:lockdown_word}}
	\caption{Word Frequencies} 
      \label{fig:words} 
    \vspace{-4mm}
\end{figure*}

\begin{figure*}[!ht]
    \centering
  \subfloat[General COVID]{%
       \includegraphics[scale=0.27]{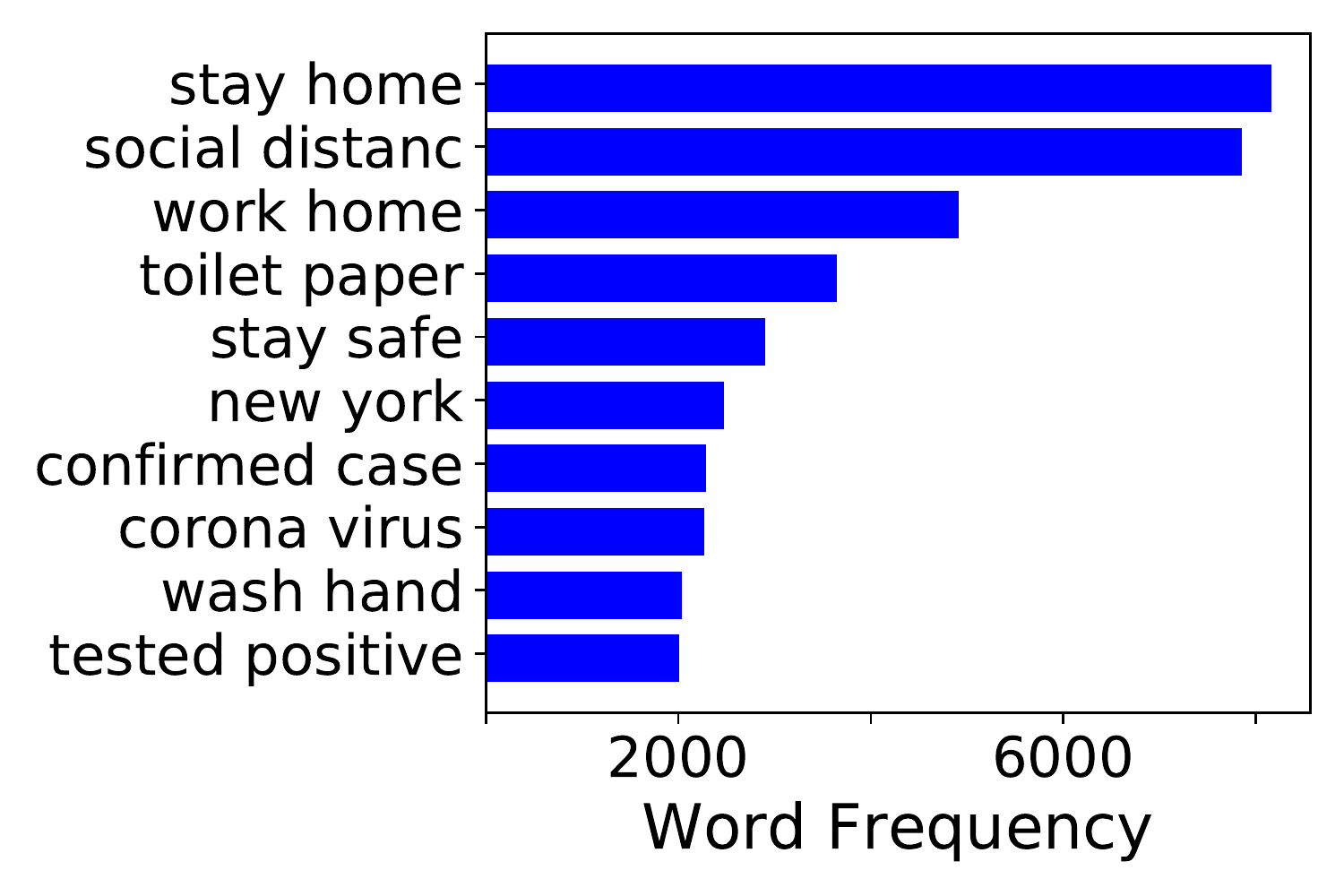}
       \label{fig:covid_word}}
       \vspace{1mm}
  \subfloat[School Closures]{%
       \includegraphics[scale=0.27]{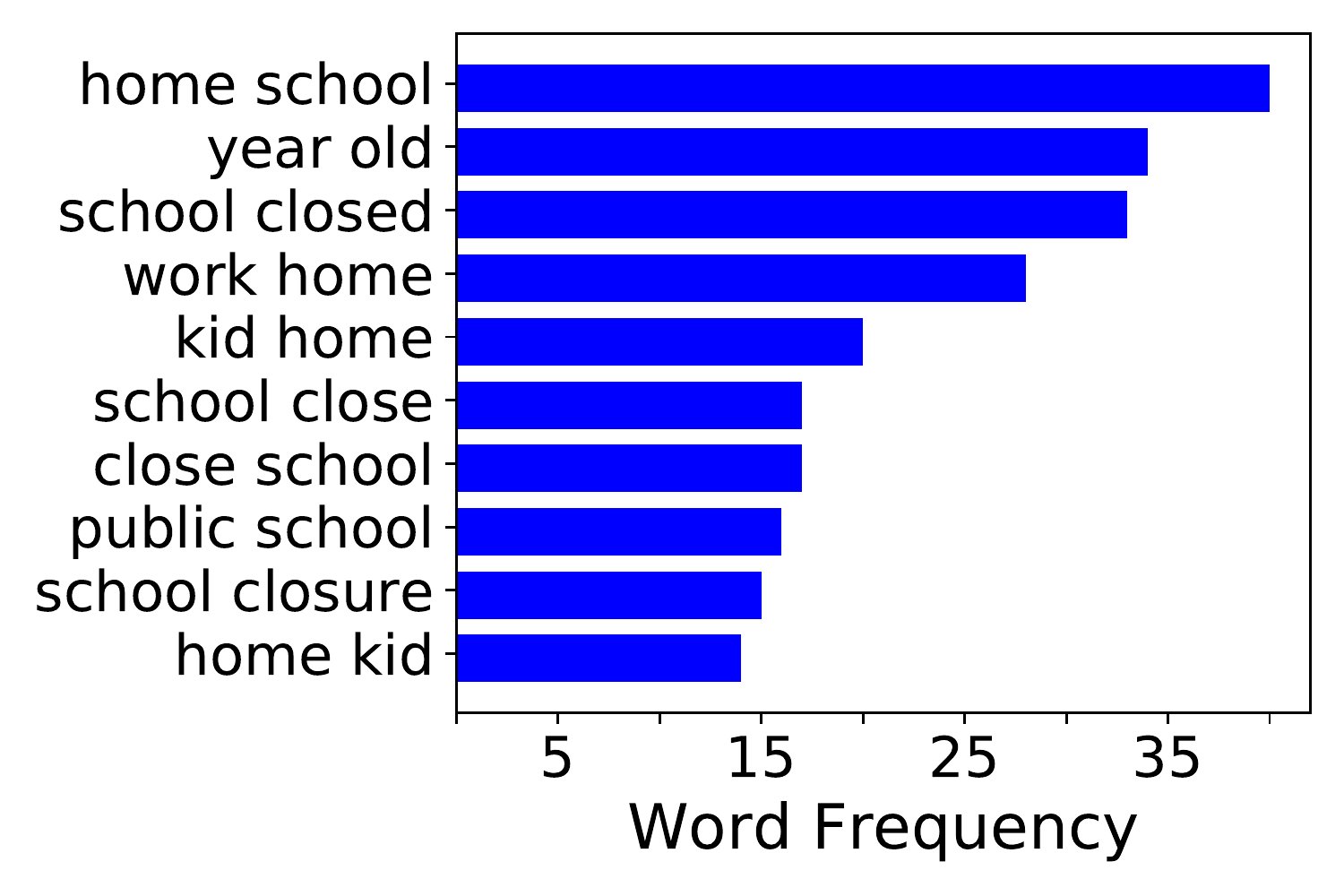}
       \label{fig:school_word}}
       \vspace{1mm}
  \subfloat[Panic Buying]{%
       \includegraphics[scale=0.27]{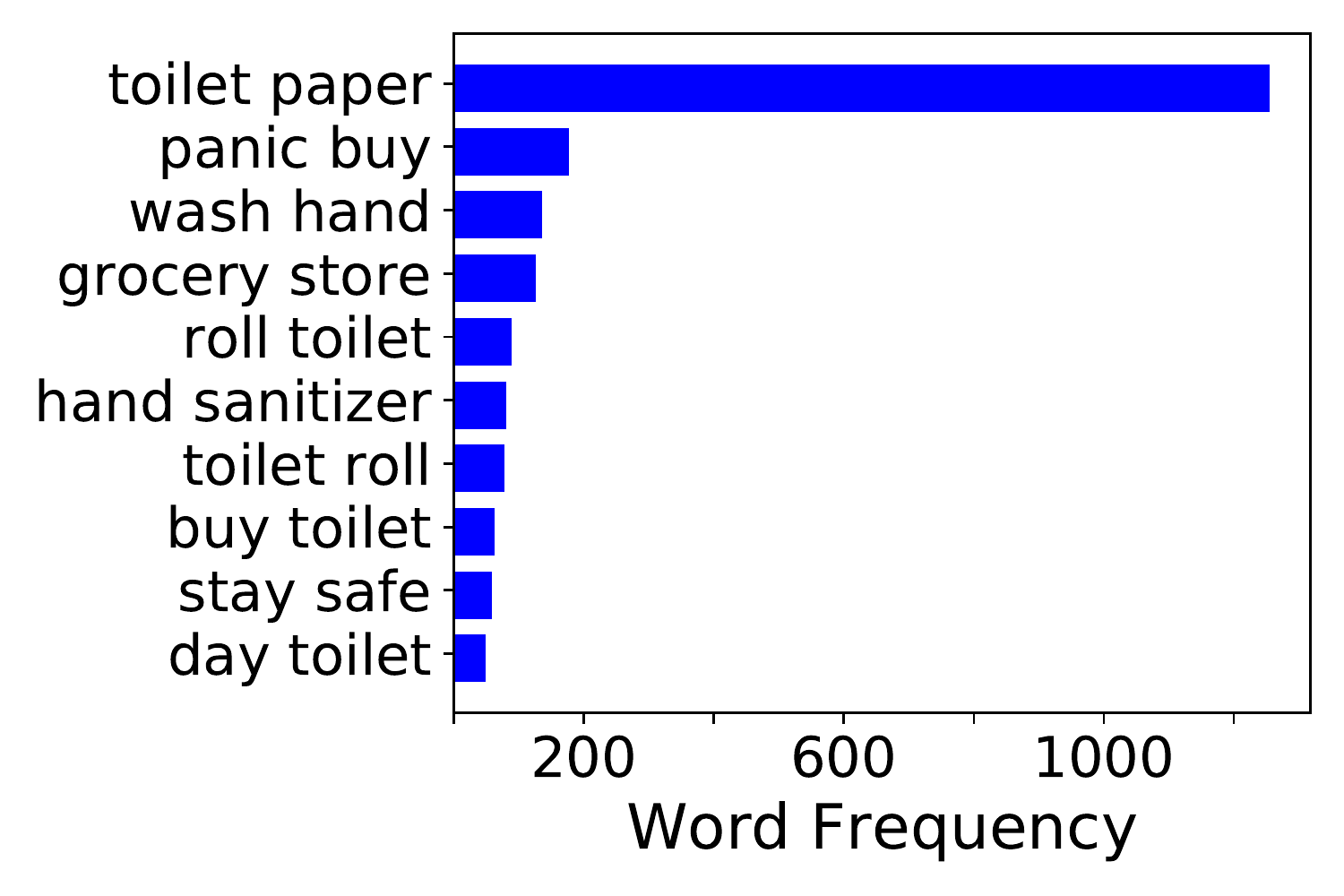}
       \label{fig:panic_word}}
       \vspace{1mm}
  \subfloat[Lockdowns]{%
       \includegraphics[scale=0.27]{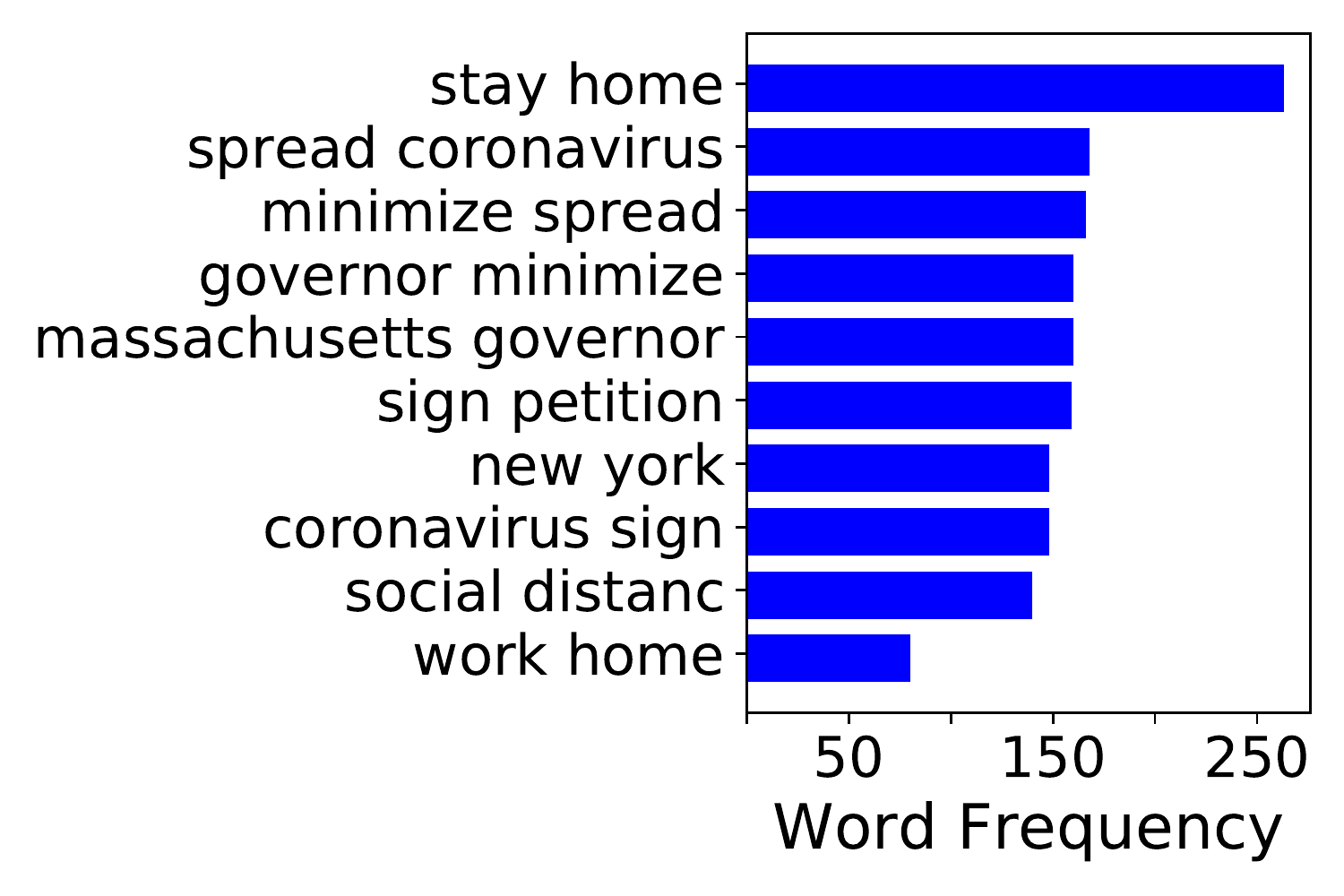}
       \label{fig:lockdown_word}}
	\caption{Bigram Frequencies} 
  \label{Fig:Bigrams} 
    \vspace{-4mm}
\end{figure*}

\subsection{Linguistic Word-Usage Analysis}
\label{sec:wordanalysis}
In this section, we present results from a linguistic word usage analysis across the different hashtag groups. Our goal is to identify the words that are uniquely representative of the particular group. To accomplish this, first, we identify and present the most commonly used words across all the hashtags. To construct the group of common words across all hashtags, we remove the words that are same or similar to the hashtags mentioned in Table \ref{tab_hashtags} as those words are redundant and tend to also be high in frequency. We also remove the names of places and governors such as New York, Massachusetts and  Andrew Cuomo. After filtering out these words, we then rank the words based on their occurrence in multiple groups and their combined frequency across all the groups. We observe words such as \textit{family}, \textit{health}, \textit{death}, \textit{life}, \textit{work}, \textit{help},  \textit{thank}, \textit{need}, \textit{time}, \textit{love}, \textit{crisis}. In Table \ref{table:examples}, we present some notable example tweets containing the common words. While one may think that \textit{health} refers to the virus-related health issues, we notice that many people also refer to \textit{mental health} in their tweets as a possible consequence to social distancing and anxiety caused by the virus. We also observe the usage of words such as \textit{death} and \textit{crisis} to indicate the seriousness of the situation. Supporting workers and showing gratitude toward them is another common tweet pattern that is worth mentioning. 
%We plan to use these observations to guide our future exploration and fine-grained analysis.

\begin{table}[ht]
\caption{Number of Tweets by Category}
\centering
\vspace{-2 mm}
\begin{tabular}{|M{7cm}|}%|M{1.8cm}|M{1.8cm}|M{1.8cm}|M{1.8cm}|M{1.8cm}|M{1.8cm}|M{1.8cm}|}
\hline
\textbf{Example of Tweets} \\
\hline
It's more important than ever to prioritize your mental \textbf{health}. You are not in this
alone.\\
 \hline
Has $13^{th}$ century returned back to $20^{th}$ century? Black \textbf{Death}. We must act very fast. \\
\hline

First, we take care of the \textbf{health} care and emergency \textbf{workers}. Then, we take care of whoever is in charge of keeping Netflix and Hulu running or it's going to get ugly \#distancesocializing \#coronavirus 
\\
 \hline
%\multicolumn{5}{l}{$^{\mathrm{a}}$Sample of a Table footnote.}
\end{tabular}
\label{table:examples}
\vspace{-3mm}
\end{table}

Second, we present the most semantically meaningful and uniquely identifying words in each hashtag group. To do this, we remove the common words calculated in the above step from each group. From the obtained list of words after the filtering, we then select the top 10 words. Due to space constraints, we only present results for four hashtag groups. Figure \ref{fig:words} gives us the uniquely identifying and semantically meaningful words in each hashtag group. In the General COVID group, we find words such as \textit{impact}, \textit{response}, \textit{resource}, and \textit{doctor}. Similarly, for School Closures, we find words such as \textit{teacher}, \textit{schedule}, \textit{educator}, \textit{book}, and \textit{class}. The Panic Buying top words mostly resonate the shortages experienced by people  such as \textit{roll} and \textit{tissue} (referring to toilet paper), \textit{hoard}, \textit{bidet} (as an alternative to toilet paper), \textit{wipe}, and \textit{water}. Top words in the Lockdown group include \textit{immigration}, \textit{shelter}, \textit{safety}, \textit{court}, and \textit{petition}, signifying the different issues surrounding lockdown.

\subsection{Word Collocation Analysis}

 We analyze words that co-occur to understand the contextual information surrounding the words. Co-occurring bigrams capture pairs of words that frequently co-occur in each group. To do this, we first filter out stop words and perform stemming and lemmatization. We calculate the overall frequencies of each word and its frequency within each class and calculate the bigram association using Pearson's Chi Squared independence test, which determines if pairs of words occur together more than they would randomly. We select the top 10 bigrams with the highest collocation statistics that are most intuitive for the human reader. Figure \ref{Fig:Bigrams} shows the top 10 bigrams for each group. We can clearly see how bigrams give better understanding compared to unigrams. Bigrams such as \textit{'toilet paper'}, \textit{'panic buy'}, \textit{'wash hand'} clearly articulate the intents of the tweets in the panic buying group. Similarly, in the lockdown group, we see \textit{'stay home'}, \textit{'work home'}, and \textit{'minimize spread'} emerging as top bigrams capturing what people are talking about in that group.

%we remove words that are already captured in the hashtags
%In Figures \ref{} we present the top occurring words in each category. To obtain these top words, we first remove the words that are already mentioned in the hashtags present in these tweets as those words are redundant and are already captured in the hashtags and tend to also be high in frequency. In the next step, we select the top 20 most occurring words in each category and then we manually select the top 10 words that are most semantically meaningful and intuitive in the group to a human reader. This step is necessary as even after removing stop words there are words such as ... that do not add much semantic value to the category.

 \subsection{Sentiment Analysis}

 To understand the sentiment across the different hashtag groups, we perform a comparative sentiment analysis. We use a pre-trained sentiment analysis model \cite{liu2019roberta}, which has 95.11\% accuracy on Stanford SST test dataset and apply it to our dataset. Our model, roBERTa base model, classifies the data into five sentiment categories: \textit{strongly positive},  \textit{positive},  \textit{neutral},  \textit{negative}, and  \textit{strongly negative}.

\begin{figure}
    \center
        \includegraphics[scale=0.55]{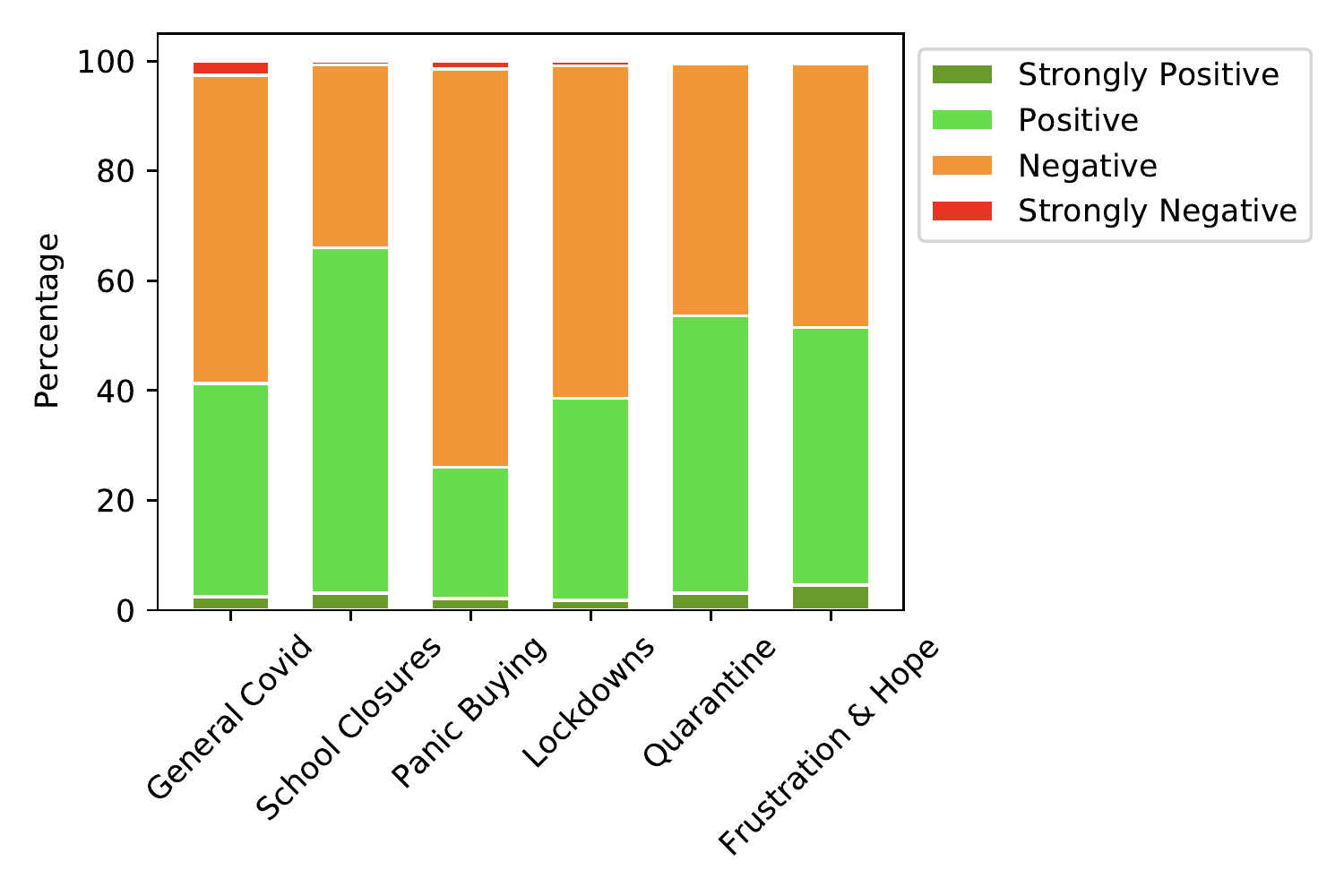}
             \vspace{-3mm}
    \caption{Multi-class Classification Sentiment Graph}
 \label{fig:sentiment}
     \vspace{-6mm}
 \end{figure}

 We present the results in Figure \ref{fig:sentiment}. Since the neutral category is not useful for our analysis, we exclude it and scale the rest of the categories to 100\%, normalizing for the number of tweets in each category. We notice that the School Closures group has a significantly higher number of positive tweets that capture the overall positive sentiment around the closure of schools. In contrast, the Panic buying group has  a higher number of negative tweets showing the frustration in relation to panic buying. Overall, we observe  \textit{strongly positive} tweets when compared to  \textit{strongly negative} tweets in all categories. This is especially interesting in the Quarantine and Frustration and Hope groups, where more tweets are showing support for quarantine and hopefulness.
 
 \begin{figure*}[!ht]
    \centering
  \subfloat[General Covid]{%
       \includegraphics[scale=0.27]{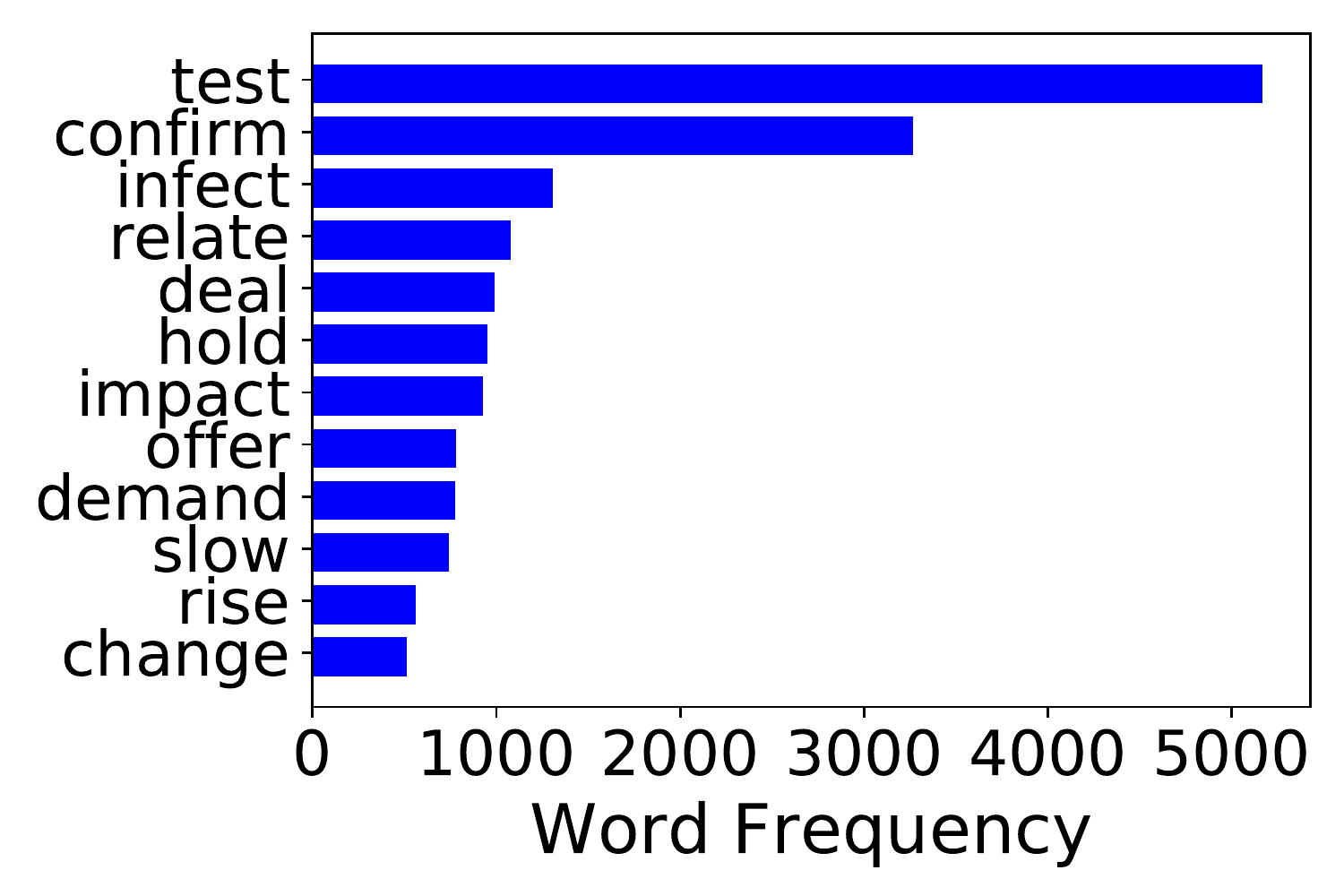}
       \label{fig:covid_word}}
       \vspace{1mm}
  \subfloat[School Closures]{%
       \includegraphics[scale=0.27]{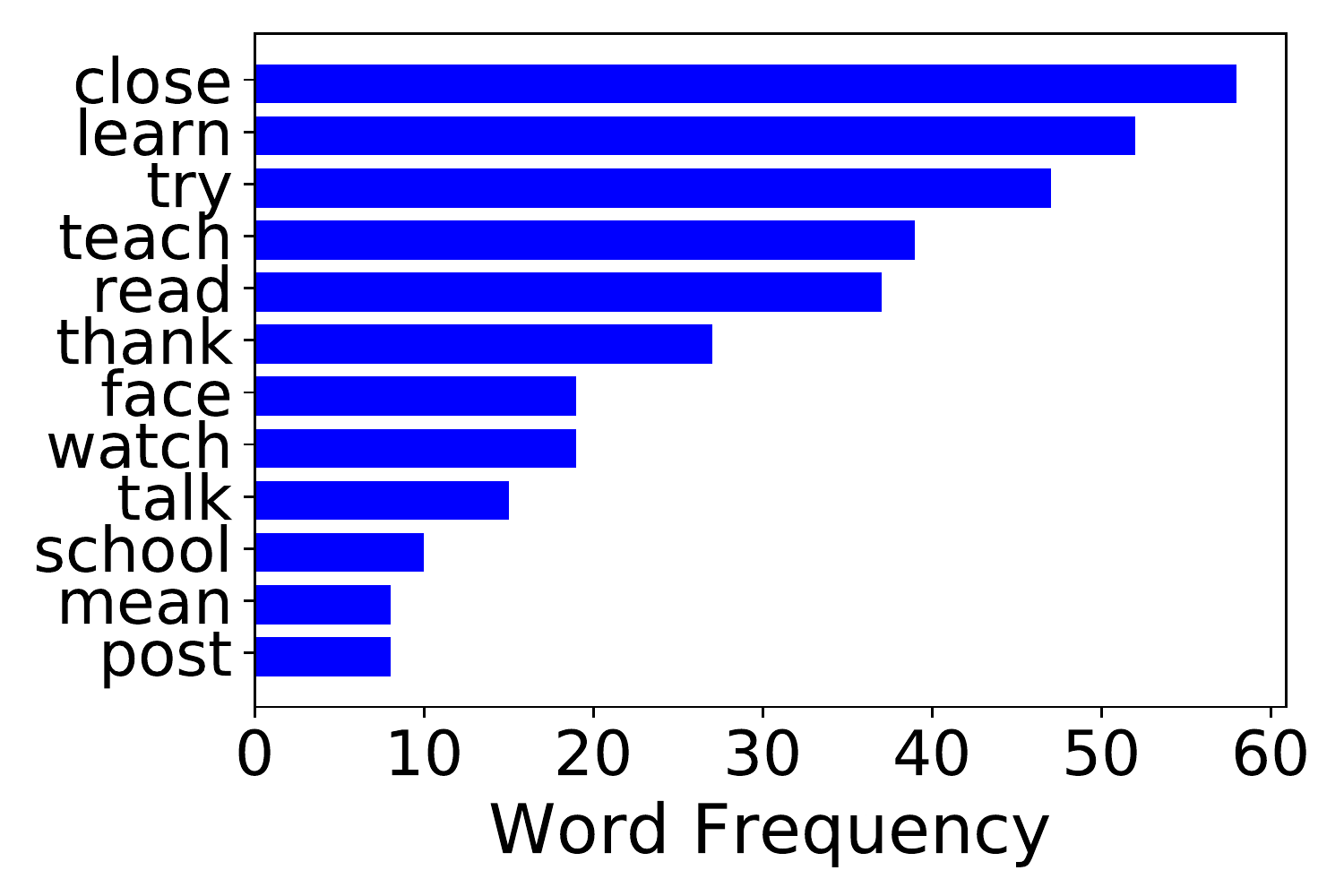}
       \label{fig:school_word}}
       \vspace{1mm}
  \subfloat[Panic Buying]{%
       \includegraphics[scale=0.27]{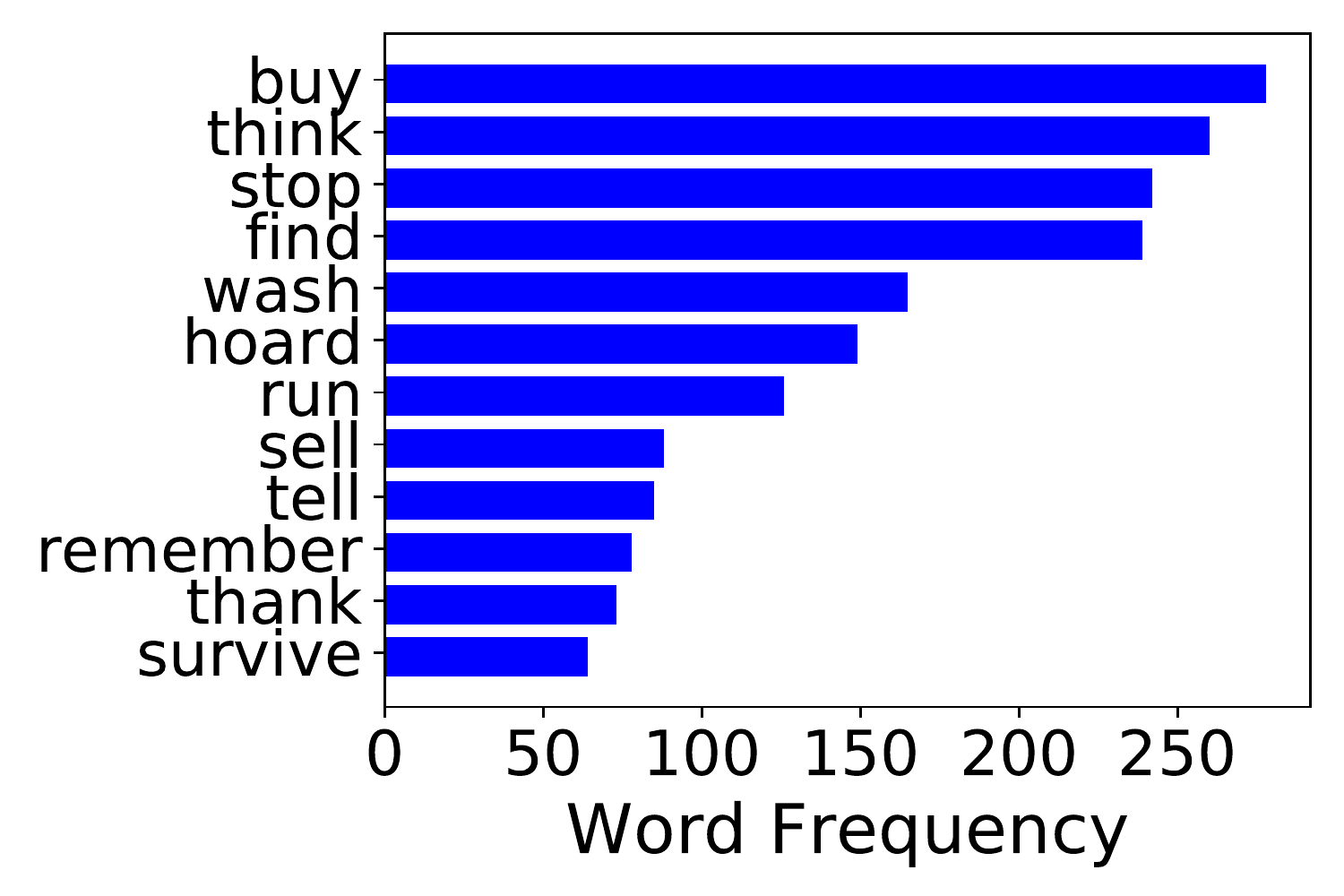}
       \label{fig:panic_word}}
       \vspace{1mm}
  \subfloat[Lockdowns]{%
       \includegraphics[scale=0.27]{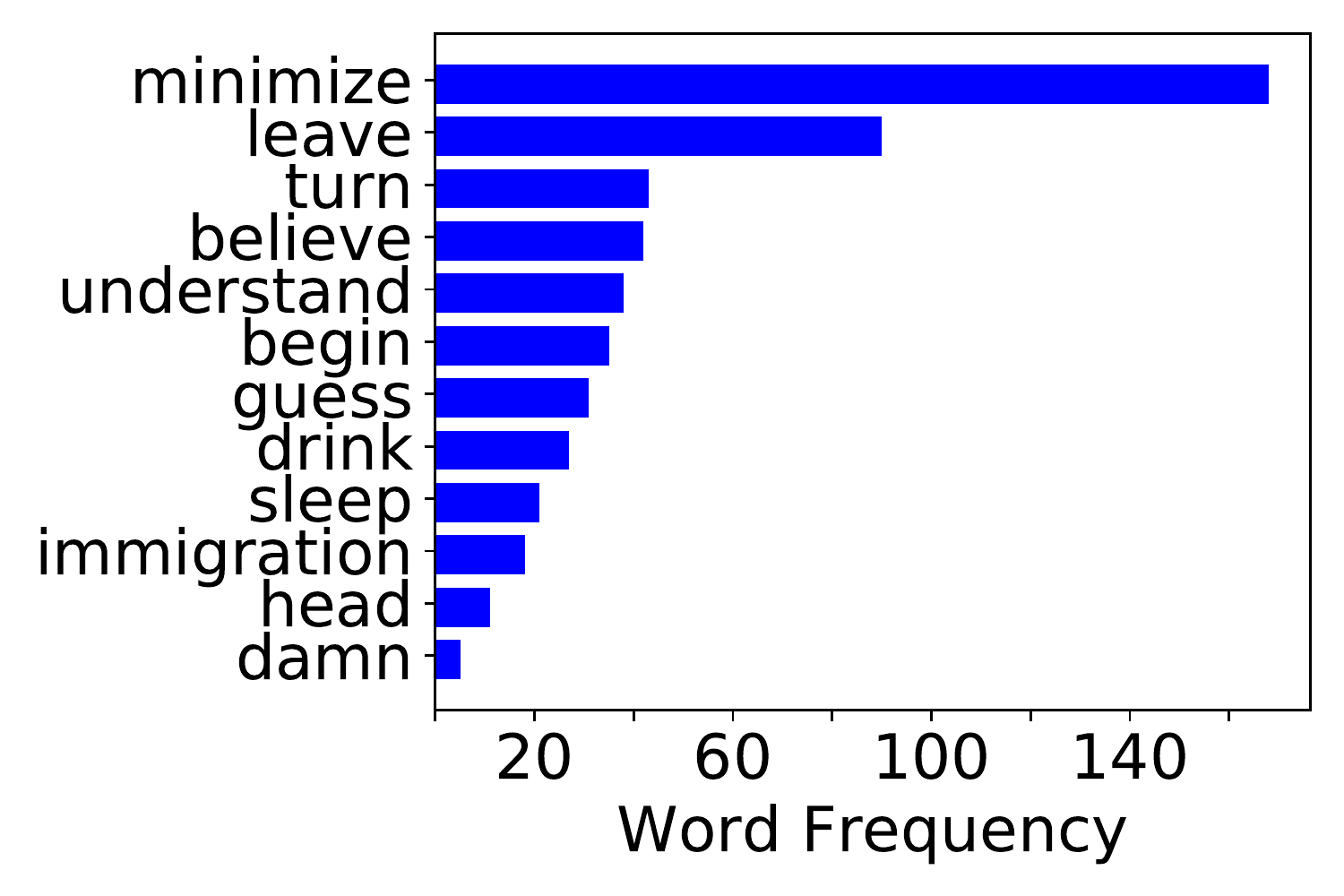}
       \label{fig:lockdown_word}}
       \vspace{1mm}
%    \subfloat[Quarantine]{%
%       \includegraphics[scale=0.28]{Figures/quarantine_group5_verb_plot.pdf}
%       \label{fig:quarantine}}
%       \vspace{1mm}
%  \subfloat[Frustration & Hope]{%
%       \includegraphics[scale=0.28]{Figures/frus_hope_xgroup6_verb_plot.pdf}
%       \label{fig:frus_hope}}
	\caption{Top Verbs in Tweets} 
  \label{fig:verbs} 
    \vspace{-4mm}
\end{figure*}

 \subsection{Semantic Role Labeling}
 \label{sec:semantic}
 We use AllenNLP BERT based model \cite{shi2019simple} to run semantic role labeling and identify the action words (verbs), which capture the actions people are referring to in the tweets. To identify the uniquely representative verbs in each group, we identify all the verbs in each group and use TF-IDF vectorization to remove the common verbs across the groups. We then compute the verb frequency of remaining verbs in each group. Figure \ref{fig:verbs} shows the verbs and their frequencies.

 The results capture the top verbs defining each group. For example, the School Closures group has \textit{close} as its top verb which signifies the closing of schools while \textit{learn}, \textit{read}, and \textit{teach} emphasize the actions corresponding to learning online because of the pandemic. In comparison some words such as \textit{mean} and \textit{post} are challenging to understand without additional context, so we present examples tweets containing these words to understand the context in which they are used in Table \ref{table:verb_examples}. All  tweets with \textit{mean} have a similar context but \textit{post} is used in two different contexts. One refers to send and the other refers to the post pandemic period.

 \begin{table}[ht]
 \center
 \caption{Tweets with \textit{Mean} and \textit{Post} verbs of School Closures Group}
 \vspace{-2 mm}
 \begin{tabular}{|M{7cm}|}%|M{1.8cm}|M{1.8cm}|M{1.8cm}|M{1.8cm}|M{1.8cm }|M{1.8cm}|M{1.8cm}|}
 \hline
 \textbf{Example of Tweets} \\
 \hline
 if they \#closetheschoolsuknow then that \textbf{means} I can self-isolate not because I’m feeling ill or showing signs of the \#coronavirus just because there's no point in going out the house if it's not for work everywhere is closed so close the schools and let me binge watch disney\\
 \hline
 I don't have many followers because I just started doing twitter again but I will \textbf{post} a hands on lesson daily for \#homeschooling  and spam all the trending hashtags. \#CoronavirusOutbreak \#StayHome \@TheTodayShow \#COVID19 \\
 \hline

 Free online classes for K-12 students from St. Louis area company Varsity tutors. St. Louis \textbf{post} dispatch reporting on the story \\
 \hline
 %\multicolumn{5}{l}{$^{\mathrm{a}}$Sample of a Table footnote.}
 \end{tabular}
 \label{table:verb_examples}
 %\vspace{-3mm}
 \end{table}
 
 Along the same lines, verbs in other groups also signify people's actions during the pandemic. The Panic Buying group captures actions such as \textit{buy}, \textit{wash}, \textit{hoard}, \textit{sell}, and \textit{survive}. Similarly, in the Quarantine group, action words such as \textit{walk}, \textit{practice}, \textit{save}, and \textit{clean} are among the top ones.

 \begin{figure}[!ht]
    \centering
  \subfloat[School Closures]{%
       \includegraphics[scale=0.40]{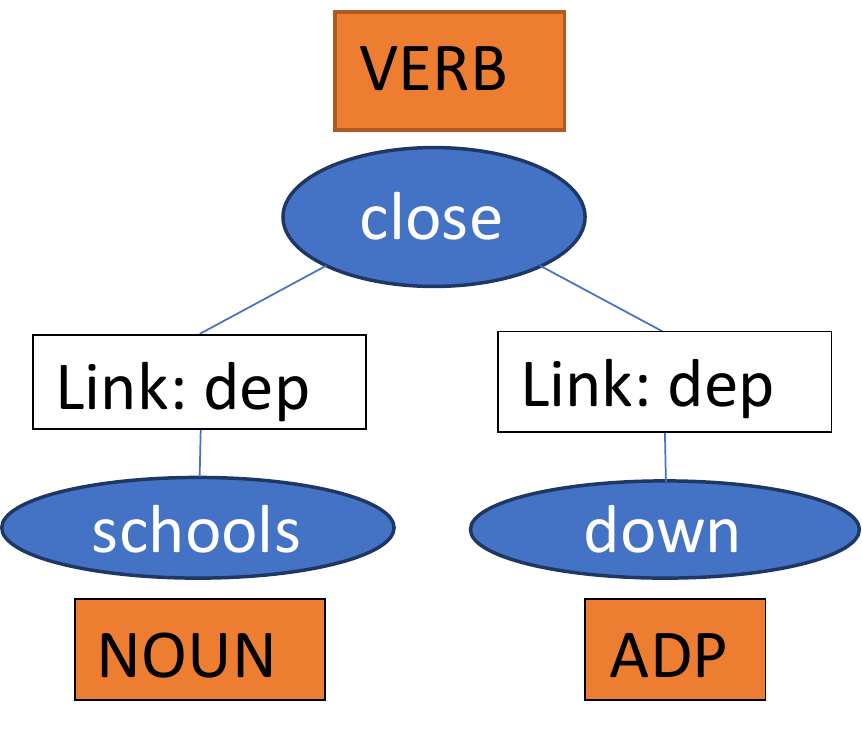}
       \label{fig:covid_graph}}
  \subfloat[Panic Buying]{%
       \includegraphics[scale=0.4]{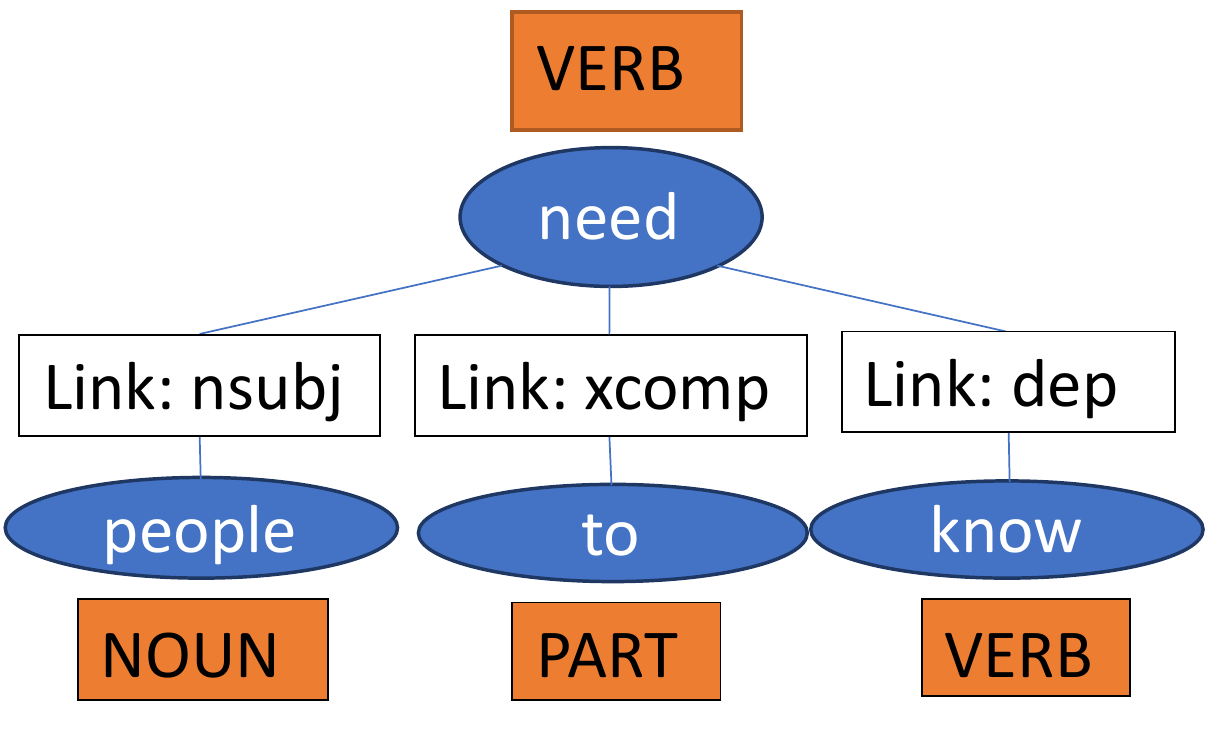}
       \label{fig:panic_graph}}
	\caption{Examples of Dependency Parse Subtrees} 
  \label{fig:tree} 
       \vspace{-6mm}
\end{figure}
 \subsection{Contextual Analysis of Action Words using Dependency Parsing}
 
To further analyze the context in which the action words discussed in Section \ref{sec:semantic} are used, we analyze the words associated with them using dependency parsing. Dependency parsing breaks down each sentence into linguistic dependency structures organized in the form of a tree. We focus on identifying the nouns that are connected to the action words. In the dependency parse, the action words/verbs form the root of the parse and the dependencies are in the left and right subtrees. To identify the nouns associated with the verbs, we traverse the dependency parse to the sub-tree where the action word of interest is present and then extract the corresponding noun. We also analyze the link associated with noun and verb and find that ``nsubj" (Nominal Subject), ``pobj" (Object of a preposition) and ``dobj" (The Direct Object) are the most related link tags that contribute to the action words. Figure \ref{fig:tree} gives some notable dependency parse subtrees with the action word and the corresponding noun. We can see that by decoding the parse structure, we can identify additional contextual information such as the nouns they refer to. 

We use the AllenNLP implementation of a neural model for dependency parsing using biaffine classifiers on top of a bidirectional LSTM \cite{dozat2016deep}. We parse the sentences associated with the top 5 verbs in Figure \ref{fig:verbs} and find their associated nouns to understand what the action verbs are used to signify. Tables \ref{table:group1_dep}, \ref{table:group2_dep}, \ref{table:group3_dep}, \ref{table:group4_dep} represent the different nouns associated with the most prominent action words in each group, respectively. General COVID being the diverse group, it contains a myriad of tweets from offering support to the fear of getting infected. Apart from the verb-noun combinations that we expect to see in the group (such as \textit{test virus}, \textit{confirm case}, \textit{offer support}), the other most notable verb-noun combinations in this group are: \textit{deal stress}, \textit{deal anxiety}, \textit{fear system}, and \textit{fear safety}. And in other groups, the verb-noun combinations narrow down on the specific actions relevant to the group. For instance, in School Closures, the action word \textit{close} mostly talks about closing the schools for benefit of students, and action word \textit{offer} co-occurs with teaching aids through online sources. In the Panic Buying group, tweets about the panic experienced by people is captured by verb-noun pairs such as \textit{stop madness}, \textit{buy paper}, \textit{find store}. In the Lockdown group, some interesting combinations surface such as \textit{believe information}, \textit{guess trust}, which captures the possible distrust people have with the lockdown measures. Nouns such as \textit{insanity} also help in capturing peoples' reaction to the lockdown measures. 
 
 \begin{table}[ht]
\caption{General Covid Action Words}
\center
\vspace{-2 mm}
\begin{tabular}{|c|c|}%|M{1.8cm}|M{1.8cm}|M{1.8cm}|M{1.8cm}|M{1.8cm}|M{1.8cm}|M{1.8cm}|}
\hline
\textbf{Action words} & \textbf{Linked Nouns} \\
\hline
test & people, president, virus, news, student \\

\hline
confirm & case, number, death, people, health \\

\hline
offer & help, advice, support, relief, delivery \\

\hline
deal & anxiety, impact, stress, virus, crisis \\

\hline
fear & people, virus, safety, system, faith \\

\hline
%\multicolumn{5}{l}{$^{\mathrm{a}}$Sample of a Table footnote.}
\end{tabular}
\label{table:group1_dep}
%\vspace{-3mm}
\end{table}

 \begin{table}[ht]
\caption{School Closures Action Words}
\center
\vspace{-2 mm}
\begin{tabular}{|c|c|}%|M{1.8cm}|M{1.8cm}|M{1.8cm}|M{1.8cm}|M{1.8cm}|M{1.8cm}|M{1.8cm}|}
\hline
\textbf{Action words} & \textbf{Linked Nouns} \\
\hline
close & school, library, mayor, classroom, times \\

\hline
learn & home, experience, life, art, material \\

\hline
thank & people, community, faculty, school, world \\

\hline
teach & matter, life, education, dad, wrong \\

\hline
mean & solidarity, educator, hunger, brain, city \\

\hline
%\multicolumn{5}{l}{$^{\mathrm{a}}$Sample of a Table footnote.}
\end{tabular}
\label{table:group2_dep}
%\vspace{-3mm}
\end{table}

 \begin{table}[ht]
\caption{Panic Buying Action Words}
\center
\vspace{-2 mm}
\begin{tabular}{|c|c|}%|M{1.8cm}|M{1.8cm}|M{1.8cm}|M{1.8cm}|M{1.8cm}|M{1.8cm}|M{1.8cm}|}
\hline
\textbf{Action words} & \textbf{Linked Nouns} \\
\hline
stop & panic, paper, madness, spread, need \\

\hline
buy & paper, toilet, stock, underwear, vaccine \\

\hline
wash & people, hand, face, plague, soap \\

\hline
find & toilet, paper, home, store, hand \\

\hline
tell & people, story, calm, deal, twitter \\

\hline
%\multicolumn{5}{l}{$^{\mathrm{a}}$Sample of a Table footnote.}
\end{tabular}
\label{table:group3_dep}
%\vspace{-3mm}
\end{table}

 \begin{table}[ht]
\caption{Lockdowns Action Words}
\center
\vspace{-2 mm}
\begin{tabular}{|c|c|}%|M{1.8cm}|M{1.8cm}|M{1.8cm}|M{1.8cm}|M{1.8cm}|M{1.8cm}|M{1.8cm}|}
\hline
\textbf{Action words} & \textbf{Linked Nouns} \\
\hline
understand & insanity, ration, diarrhea, crisis, situation \\

\hline
turn & chapter, point, downtown, person, thing \\

\hline
believe & information, lesson, territory, reason, motto \\

\hline
guess & target, people, defecation, house, trust \\

\hline
begin & voilence, shop, problem, news \\

\hline
%\multicolumn{5}{l}{$^{\mathrm{a}}$Sample of a Table footnote.}
\end{tabular}
\label{table:group4_dep}
%\vspace{-3mm}
\end{table}

\subsection{Seeded Topic Models}

In this section, we use Seeded LDA \cite{jagarlamudi2012incorporating}  to categorize the tweets and check the closeness of these automatically obtained groups  with our manual grouping using the hashtags. As we are specifically interested in isolating the tweets in specific topics of our interest than general topics identified by a topic model, we leverage a seeded variant of LDA, Seeded LDA \cite{jagarlamudi2012incorporating} to guide the topic model to discover them. Seeded LDA allows seeding of topics by providing a small set of keywords to guide topic discovery influencing both the document-topic and the topic-word distributions. The seed words need not be exhaustive as the model is able to detect other words in the same category via co-occurrence in the dataset. Our goal with seeded LDA is to i) present a way to automatically categorize tweets into specific topics of interest, especially when the topics are rarer in the dataset, ii) passively evaluate the effectiveness of our word analysis thus far, and iii) develop a scalable approach that can be extended to millions of tweets with minimal manual intervention. 

We develop a Seeded LDA model to categorize tweets into the five hashtag groups: i) General COVID, ii) School Closures, iii) Panic Buying, iv) Lockdowns, and  v) Quarantine by seeding each group with seed words from our analysis in Section \ref{sec:wordanalysis}. We leave out the Frustration and Hope topic due to the inherent polarizing nature of the keywords and the lack of identifying keywords that are unique for the topic. We select the top few words from our words in Figure \ref{fig:words} as seed words for our Seeded LDA model. Table \ref{table:seeds} gives the seed words for the different COVID categories. We include $k$ un-seeded topics in our model to account for messages that do not fall into these topic categories. After experimenting with different values of $k$ and manually evaluating the topics, we find that $k$ = 2 gives us the best separation and categorization. We use $\alpha$ = 0.01 and $\beta$ = 0.0001 to give us sparse document-topic and topic-word distributions where fewer topics and words with high values emerge, so we can classify the tweets to the predominant category. We train the seeded LDA models for 2000 iterations. We first use the document-topic distribution to get the best topic for each tweet. If the best topic of the message is one of the seeded topics which correspond to the categories, then, we classify the tweet into that category. In the event that a clear best topic does not emerge, we randomly assign the tweet to one of the topics that have the same document topic distribution.

\begin{table}[ht]
\caption{Seed List for Topic Modeling}
\center
\vspace{-2 mm}
\begin{tabular}{|c|c|}%|M{1.8cm}|M{1.8cm}|M{1.8cm}|M{1.8cm}|M{1.8cm}|M{1.8cm}|M{1.8cm}|}
\hline
\textbf{Category} & \textbf{Seed words} \\
\hline
General COVID & pandemic,test,covid,spread \\

\hline
School Closures & school,teach,read,student,class \\

\hline
Panic Buying & roll,shop,panic,toilet,paper,buy,sanitize \\

\hline
Lockdowns & city,shutdown,shut,state,close,petition \\

\hline
Quarantine & quarantine,work,stay,social,distance,home \\

\hline
%\multicolumn{5}{l}{$^{\mathrm{a}}$Sample of a Table footnote.}
\end{tabular}
\label{table:seeds}
%\vspace{-3mm}
\end{table}

\subsubsection{Analyzing Effectiveness of Seeded LDA Model}

To check how closely the hashtag groups match with the seeded LDA groups, we  measure the accuracy by comparing the document topic distribution from the LDA against the grouping determined by the hashtags. We do this by calculating the confusion matrix which gives us four metrics such as true positives, true negatives, false positives, and false negatives to further calculate accuracy, precision, recall, and f1 scores, which gives the value of correctness. The results we obtained are shown in the Table \ref{table:correctness}. This endeavor helps in determining the effectiveness of our word analysis (seeds) and our seeded LDA model. 

\begin{table}[ht]
\center
\caption{Model Correctness results}
\vspace{-2 mm}
\begin{tabular}{|c|c|}%|M{1.8cm}|M{1.8cm}|M{1.8cm}|M{1.8cm}|M{1.8cm}|M{1.8cm}|M{1.8cm}|}
\hline
\textbf{Metric} & \textbf{Result(in Percentage)} \\
\hline
Accuracy & 69.10852 \\

\hline
Precision & 41.04047 \\

\hline
Recall & 38.84685 \\

\hline
F1 Score & 39.91354 \\
\hline
%\multicolumn{5}{l}{$^{\mathrm{a}}$Sample of a Table footnote.}
\end{tabular}
\label{table:correctness}
%\vspace{-3mm}
\end{table}

Also, to verify that our model had best results in the groups that we are interested in, we calculate the precision, recall, and F1 scores for the School Closures, Panic Buying, Lockdowns, and Quarantine groups. We exclude General COVID and Frustration and Hope groups as they are too general, and we are interested in isolating the more specific COVID groups. Table \ref{table:groups} shows the results of each group.

\begin{table}[ht]
\center
\caption{Model correctness on Individual groups}
\vspace{-2 mm}
\begin{tabular}{|c|c|c|c|}%|M{1.8cm}|M{1.8cm}|M{1.8cm}|M{1.8cm}|M{1.8cm}|M{1.8cm}|M{1.8cm}|}
\hline
\textbf{Groups} & \textbf{Recall} & \textbf{Precision} & \textbf{F1 Score} \\
\hline
School Closures & 52.38567 & 30.37499 & 38.45340 \\

\hline
Panic Buying & 52.09513 & 30.58002 & 38.53806 \\

\hline
Lockdowns & 21.17116 & 34.11569 & 26.12805 \\

\hline
Quarantine & 24.17547 & 88.51389 & 37.97811 \\

\hline
%\multicolumn{5}{l}{$^{\mathrm{a}}$Sample of a Table footnote.}
\end{tabular}
\label{table:groups}
%\vspace{-3mm}
\end{table}

By examining the result, we observe that the manual grouping of the hashtags have significant match with the seeded LDA groups. We also note that the seeded LDA model is able to correctly isolate the tweets in rarer groups where there is less data, such the School closures group. This shows the effectiveness of our model to analyze rarer groups in the data. Additionally, from the precision of classification for the Quarantine group, we observe that the false positives were significantly low and further adds credibility to our model.

\section{Related Work}
\label{sec:related}
%!TEX root = main.tex

In this section, we  outline existing research related to modeling and analyzing Twitter and web data to understand social, political, psychological, and economic impacts of a variety of different events. Due to the recent nature of the outbreak, there is little to no published work on COVID-19. We primarily focus on discussing work that analyze Twitter communications. Ahmed et al. focus on the conspiracy theories surrounding the novel coronavirus, especially in relationship with 5G \cite{ahmed2020covid}. The authors analyze Twitter communications and discuss the possibility of using bots for propagating misinformation and political conspiracies during the pandemic \cite{kouzy2020coronavirus},  \cite{Ferrara_2020}.  In comparison,  the authors in \cite{park2020conversations} conduct infodemiology studies on Twitter communications to understand how information is spreading during this time, while the  the stigma created by referencing the novel coronavirus as ``Chinese virus'' is investigated in \cite{budhwani2020creating}. Twitter has been used to study political events and related stance \cite{JG_coling_2016,le2017bumps}, human trafficking \cite{tomkins2018impact}, and public health \cite{perez2019using,Dredze:IS12,Balani:DCM15,Choudhury:AAAI13,McIver:jmir15}. Several work perform fine-grained linguistic analysis on social media data \cite{zhang2018fine,tomkins2018socio,raisi2018weakly}.

%po Tomkins et al. analyze the effect of environmental stressors on human trafficking. Zhang et al. perform a linguistic analysis of an anonymous messaging application. 

\section{Discussion and Concluding Remarks}
\label{sec:conclusion}
%!TEX root = main.tex

In this paper, we studied Twitter communications in the United States during the early days of the COVID-19 outbreak.  As the disease continued to spread, we observed  panic buying as well as  calls for closures of schools, bars, cities, social distancing and quarantining.  We  conducted a linguistic word-usage analysis and identified the most frequently occurring unigrams and bigrams in each group that provide us an idea of the main discussion points.   We conducted sentiment analysis to understand the extent of positive and negative sentiments in the tweets. We then performed semantic role labeling to identify the key action words and then obtained the corresponding contextual words using dependency parsing. Finally, we designed a scalable seeded topic modeling approach to automatically identify the key topics in the tweets.

%Also, our qualitative analysis reveals that the words are used in multiple different contexts, which is worth delving further into. T
%and its primary aim is to quantitatively outline the socio-economic distress already caused by COVID-19 so that we as a society can learn from this experience and be better prepared if COVID-19 (or maybe another pandemic) were to (re)emerge in the future.  At the time of writing this paper, the infection spread has still not reached its peak in the United States. Therefore, we plan to keep collecting data to understand and investigate the socio-economic and political impact of COVID-19. 

%\bibliographystyle{ACM-Reference-Format}
%\bibliography{references}

\bibliographystyle{IEEETran}
\bibliography{references}

\end{document}